\documentclass[fleqn,usenatbib]{mnras}

\usepackage{newtxtext,newtxmath}
\usepackage{siunitx}

\usepackage[T1]{fontenc}

\DeclareRobustCommand{\VAN}[3]{#2}
\let\VANthebibliography\thebibliography
\def\thebibliography{\DeclareRobustCommand{\VAN}[3]{##3}\VANthebibliography}

\usepackage{graphicx}	
\usepackage{amsmath}	

\newcommand{\mmp}{$\pm$}

\newcommand{\nm}{\,nm}
\newcommand{\dst}{\,kg\,m$^{-3}$}
\newcommand{\mgs}{\,mg\,s$^{-1}$}
\newcommand{\ti}{\,J\,K$^{-1}$\,m$^{-2}$\,s$^{-1/2}$}
\newcommand{\tc}{\,W\,m$^{-1}$\,K$^{-1}$}
\newcommand{\mc}{\,$\umu$m }

\usepackage[table]{xcolor}
\usepackage{makecell}
\definecolor{blu}{RGB}{0,158,115}
\definecolor{ver}{RGB}{213,94,0}
\definecolor{rei}{RGB}{204,121,167}

\usepackage{colortbl}
\usepackage{arydshln}

\title[Cometary dust analogues for physics experiments]{Cometary dust analogues for physics experiments}

\author[A. Lethuillier et al.]{A. Lethuillier,$^{1}$\thanks{E-mail: a.lethuillier@tu-bs.de}
C. Feller,$^{3}$
E. Kaufmann,$^{2,7}$
P. Becerra,$^{3}$
N. Hänni,$^{3}$
R. Diethelm,$^{1}$
C. Kreuzig,$^{1}$
\newauthor
B. Gundlach,$^{1}$
J. Blum,$^{1}$
A. Pommerol,$^{3}$
G. Kargl,$^{2}$
E. Kührt,$^{4}$
H. Capelo,$^{3}$
D. Haack,$^{4}$
X. Zhang,$^{6}$
\newauthor
J. Knollenberg,$^{4}$
N. S. Molinski,$^{1}$
T. Gilke,$^{1}$
H. Sierks,$^{5}$
P. Tiefenbacher,$^{2}$
C. Güttler,$^{5}$
K. A. Otto,$^{4}$
\newauthor
D. Bischoff,$^{1}$
M. Schweighart,$^{2}$
A. Hagermann,$^{7}$
N. Jäggi,$^{3}$
\\
$^{1}$Institut für Geophysik und extraterrestrische Physik (IGeP), TU Braunschweig, Mendelssohnstr. 3, 38106 Braunschweig, Germany.\\
$^{2}$Space Research Institute, Austrian Academy of Science, Schmiedlstraße 6, 8042 Graz, Austria.\\
$^{3}$Physikalisches Institut, Universität Bern, Sidlerstrasse 5, 3012 Bern, Switzerland.\\
$^{4}$Deutsches Zentrum für Luft- und Raumfahrt, Rutherfordstraße 2, 12489 Berlin-Adlershof, Germany.\\
$^{5}$Max-Planck-Institut für Sonnensystemforschung, Justus-von-Liebig-Weg 3, 37077 Göttingen, Germany.\\
$^{6}$Qian Xuesen Laboratory of Space Technology, China Academy of Space Technology.\\
$^{7}$Department of Computer Science, Electrical and Space Engineering, Lulea University of Technology, Bengt Hultqvistsvag 1. 981 92 Kiruna, Sweden\\
}

\date{Accepted XXX. Received YYY; in original form ZZZ}

\pubyear{2020}

\begin{document}
\label{firstpage}
\pagerange{\pageref{firstpage}--\pageref{lastpage}}
\maketitle

\begin{abstract}
The CoPhyLab (Cometary Physics Laboratory) project is designed to study the physics of comets through a series of earth-based experiments. For these experiments, a dust analogue was created with physical properties comparable to those of the non-volatile dust found on comets. This "CoPhyLab dust" is planned to be mixed with water and CO$_2$ ice and placed under cometary conditions in vacuum chambers to study the physical processes taking place on the nuclei of comets. In order to develop this dust analogue, we mixed two components representative for the non-volatile materials present in cometary nuclei. We chose silica dust as representative for the mineral phase and charcoal for the organic phase, which also acts as a darkening agent. In this paper, we provide an overview of known cometary analogues before presenting measurements of eight physical properties of different mixtures of the two materials and a comparison of these measurements with known cometary values. The physical properties of interest are: particle size, density, gas permeability, spectrophotometry, mechanical, thermal and electrical properties. We found that the analogue dust that matches the highest number of physical properties of cometary materials consists of a mixture of either 60\%/40\% or 70\%/30\% of silica dust/charcoal by mass. These best-fit dust analogue will be used in future CoPhyLab experiments.

\end{abstract}




\section{Introduction}

The study of the physics of planetary analogue materials is essential for our understanding of the objects of the Solar System. In particular, research on the physics of cometary materials greatly benefits from laboratory experiments, as in-situ measurements are difficult to obtain. Our knowledge of comets and cometary nuclei has increased significantly in the last few decades. With the help of numerical models, earth-based observations, and especially ambitious space missions, a remarkable amount of information has been collected, providing insights into the physical processes taking place on comets.

The Rosetta mission in particular, spent more than two years studying comet 67P/Churyumov-Gerasimenko (hereafter 67P/C-G) and its nucleus from up close with a wide array of instruments. In November 2014, the Philae lander detached from the Rosetta orbiter and descended to land on the surface of the nucleus. For three days, the lander measured many physical properties of the surface before running out of power.

In order to interpret many of the measurements performed by the Rosetta instruments, models had to be developed and assumptions on the materials that make up the nucleus had to be made. Experimental studies of cometary analogues in environments similar to those encountered in comets can help to validate the models and limit the assumptions made about the composition of comets. Support from laboratory experiments to cometary science has been ongoing since the 1970's. The first cometary experiment was done by \cite{dobrovolsky_kajmakov_1977} who studied the sublimation of water ice/refractory mixtures in vacuum. Similar experiments were conducted at the Jet Propulsion Laboratory in the 1980's by \cite{STEPHENSAUNDERS1986} and \cite{STORRS1988}. One of the most ambitious cometary laboratory experiments was the KOSI ("Kometen-SImulation", Comet-Simulation in English) experiments by \cite{Grun1992}, in which mixtures of dust and ice were cooled down to cryogenic temperatures before being illuminated in vacuum. These experiments allowed the measurements of the thermo-mechanical properties of cometary analogues and also showed that thermal activity within the samples led to the formation of a desiccated refractory layer with a low thermal conductivity \citep{Spohn1989}. However, these experiments had many free parameters and the properties changed between measurements, which complicated the interpretation of some of the observations.

The data sent back by Rosetta from comet 67P/C-G provided us with an unprecedented view of a comet and has given us access to the physical properties of cometary material. Thanks to these recent measurements made by all of the Rosetta instruments, it is now possible to create more accurate refractory cometary dust analogues. These dust analogues will need to be mixed with volatile components in order for earth-based experiments to shed light on the physical processes happening on comets.

To this end, in the framework of the CoPhyLab (Comet Physics Laboratory, \citealt{CoPhyLab2020}) project, we present our cometary dust recipe. The main goal of this refractory analogue is to reproduce the physical properties of actual cometary material as accurately as possible while being procurable in large enough amount for laboratory experiments. The sample should also be representative of different comets, not just 67P/C-G. As the CoPhyLab experiments are focused on physical and not chemical properties, we do not aim to reproduce the chemical composition of comets with this analogue. We acknowledge that chemical and physical composition are interdependent of each other and at a later stage, to study more complex processes, a chemical analogue might be desirable but is out of the scope of this paper. 

In Section~\ref{KnowCOmetaryAnalogs}, we will present a brief overview of current cometary analogues, we will also justify the development of the CoPhyLab comet refractory analogue (hereafter CoPhyLab dust) and specifically the use of a two component mixture of silica and charcoal. In Section \ref{sec:materialproperties}, we put forward a detailed list of all the physical properties of interest, a justification for their selection, and the target values for these properties. Finally, in Section \ref{sec:cophylabdust}, we document the procedure we use to mix the CoPhyLab dust, as well as the physical properties of different mixing ratios of silica dust and charcoal.

\section{Known cometary analogues}
\label{KnowCOmetaryAnalogs}
A few groups have developed and used cometary analogues in the past, with the goal of studying the sublimation of volatiles from an illuminated dust/ice mixture. These cometary analogues have a lot in common but also have some significant particularities that depend mostly on the experiment they were used in as we discuss below.

For the KOSI experiment \citep{Grun1992}, large quantities of samples of up to a few kg had to be produced. Therefore, the material used as an analogue of the refractory part had to be readily available. \citet{Stoffler1991} mention the use of mixtures of silicate minerals and soot as the refractory parts in a number of KOSI experiments. The silicate component was a mixture of olivine, pyroxene, montmorillonite, and kaolinite originating from terrestrial sources. They were chosen following observations made on comet Halley, interplanetary dust particles, and carbonaceous chrondrites \citep{Stoffler1991}. The soot was chosen as a carbon source and to control the albedo of the samples. In total, 6 samples were created each one for one of the KOSI experiments \citep{Grun1991}. Later experiments used olivine mixed with soot for more complex sublimation experiments \citep{Grun1993}. Similar cometary analogues were used in later studies, such as in \cite{HERIQUE2002}, in which the dielectric properties of samples at radar frequencies were investigated in preparation for the Rosetta mission.

Other mixtures of silicate minerals and carbon were analysed in preparation for the arrival of Rosetta and for the analysis of cosmic dust measurements. \cite{ROTUNDI2002} studied different mixtures of Mg and Fe bearing silicates, carbon rich dust, and mixed carbon silicate dust. The samples were then exposed to extraplanetary post-condensation processes, analysed with electron microscopy, and spectroscopy to identify their textures, morphologies, grain compositions, and crystallographic properties.

The KOSI samples were relatively complex, as they were made up of multiple ground minerals mixed with soot. This could result in difficulties when analysing various measurements quantitatively. More recent studies have tended towards the use of simpler mixtures, e.g., \cite{Kaufmann2017} used carbon black as the refractory part of the cometary material, which served mainly as a light absorbing agent to induce morphology changes under illumination conditions. For the specific measurements presented in their study, such as hardness measurements and morphology changes, this mixture was sufficient. In \cite{FLANDES2018}, aerogel was used as a cometary dust analogue to study the trajectory of a dust particle. Therefore, only the low bulk density and Young's modulus needed to be reproduced by the analogue.

Nevertheless, for some measurements more complex samples are necessary. For example, in an effort to reproduce the spectroscopic and the photometric properties of the nucleus' surface of comet 67P/C-G, \citet{Poch_2016,Poch_2016a} and \citet{Jost_2017a,JOST2017} produced their own cometary analogue. The dust component of this analogue were made of tholins as analogues of the complex organic matter needed to reproduce the strong spectral slope in the visible and carbon powder to control the albedo of the sample. Noting that the presence of silicate materials appears to have little effect on the spectral properties of such mixtures \citep{ROUSSEAU2018}, silicates were not included in these studies, in order to lower the number of components and make the observations easier to interpret.

The CoPhyLab project comprises a great deal of experimental facilities dedicated to investigating the physics of comets \citep{2019Kreuzig,2021Kreuzig}. Experiments are planned to study the sublimation of ices, ejection of particles, heat transfer inside the surface layers, gas diffusion through the material, mechanical properties, electrical properties, morphology, and evolution of spectrophotometric properties \citep{CoPhyLab2020}. Therefore, the dust analogues previously described are not suitable, as they generally are aimed at studying one physical parameter in particular. Also, the complex analogues used in the KOSI experiments are not suitable, because they contain too many components with too many unknown physical properties, which would make the interpretation of the measurements more difficult. In order for these experiments to correctly reflect the physical processes occurring on comets, a nucleus refractory analogue must be created and well characterised. The properties that the CoPhyLab project is interested in studying are physical and therefore we do not require a chemical match to the estimated cometary composition. However, we note that the chemical composition has a strong effect on key physical properties such as albedo/density and therefore cannot completely be ignored. In the following Section, we describe each physical parameter of interest, the values we are aiming for with our dust analogue, and our methods of measurement.

\section{Material properties}
\label{sec:materialproperties}

The samples studied in the CoPhyLab large chamber weigh several kg  \citep{2019Kreuzig,2021Kreuzig}, therefore the comet analogue has to consist of safe components purchasable in quantities in the kg range for an affordable price. Our dust analogue is thus composed of a two phase mixture:
\begin{itemize}
  \item The first phase is made of silicon dioxide (S5631), otherwise known as silica, from Honeywell Research Chemicals. This component simulates the silicate materials. It comes in the form a white dust with a well controlled particle size (99\,\% 0.5-10\,\micron) and can be purchased in 1 kg increments at a price of currently $\sim$200\,€\,Kg$^{-1}$).
  \item The second phase is Juniper charcoal dust purchased from (Pyropowders Wacholderkohle). This component was chosen to represent the organic matter known to be present on the surface of the comet, but also to act as a darkening agent. It comes in the form of a dark powder with a larger particle size than the silicon dioxide (<100\,\micron) and a lower cost (currently 12\,€\,Kg$^{-1}$). Even if charcoal is known to not be present on comets, it is generally used as an analogue for the dark organic material present on the surface as will be described later. We will also see that it allows us to match many of the physical properties of cometary dust. We were not able to find charcoal with a similar particle size to the silica dust in the quantities necessary for the CoPhyLab experiments. Grinding the charcoal, in the quantities needed for the experiments, is feasible but would require a lot of work-power and time. As the non ground charcoal allows us to fit many physical parameters, grinding was not deemed necessary. 
\end{itemize}

A total of eight physical properties were deemed essential for the CoPhyLab physics experiments. For each property, a short description of the estimated cometary values and how we measure them in our laboratory is presented below.

\subsection{Particle size distribution}

\subsubsection{Cometary particle size distributions}
\label{sec:cometarydustvalues}
The state of knowledge on the particle size distribution of cometary refractories has greatly increased since Rosetta. Previous studies of chondritic porous interplanetary dust particles from the Earth’s stratosphere have shown particles with sub-micrometre sizes that generally follow log-normal size-frequency distributions \citep{RIETMEIJER1993,2013Wozniakiewicz}. Studies of the particles collected by the Stardust mission also show a strong presence of particles with sizes below the micrometre range that follow log-normal distributions \citep{Horz2006, Zolensky2006}.

Rosetta offered many opportunities to constrain the particle size distribution of cometary dust. Visible and Infrared Thermal Imaging Spectrometer (VIRTIS) observations of the coma and of outbursts have shown the presence of sub-micrometre particles and a size distribution with a power-law index in the range 2.5–3 \citep{Bockel_e_Morvan_2017}. In addition, an anomalous feature observed by the Alice spectrograph can be explained by the presence of nanometre grains close to the instrument \citep{NOONAN2016}. \cite{Johansson2017} observed a strong decrease in the photoelectric current measured by the Langmuir Probe at perihelion that could be linked to the presence of nanometre-sized grains.

Three instruments on board Rosetta were dedicated to the study of the cometary dust: MIDAS (Micro-Imaging Dust Analysis System), GIADA (Grain Impact Analyser and Dust Accumulator), and COSIMA (Cometary Secondary Ion Mass Analyser). MIDAS is the instrument with the best size resolution and has allowed for a classification of the collected dust particles in the coma into three categories \citep{Mannel2019}:
\begin{enumerate}
    \item Large agglomerates ($\approx$10\,\micron) composed of micrometre sized particles in a fragile arrangement.
    \item Large agglomerates ($\approx$10\,\micron) composed of micrometre sized particles in a structure with fractal dimension less than two.
    \item Small particles ($\approx$1\,\micron) with subunits in the 100\,nm range.
\end{enumerate}
In addition, \cite{Mannel2019} also found that the size distribution of the smallest particles follow a log-normal law with an average around 100\,nm and a power law index in the range 2.9 to 3.8.

The analysis of the COSIMA data has revealed the presence of agglomerates in the range of 30 to 150\,\micron~\citep{Merouane2017} and that the power law index of the particle size distribution changes as the comet orbits the sun (from -1.6 after equinox to -2.8 at perihelion, \citealt{Merouane2017}). The lack of small particles detected (below 30\,\micron) is interpreted as the fact that small charged particles would have been deflected before reaching the instrument. Large agglomerates (>500\,\micron) were also detected but they disappeared after equinox \citep{Merouane2017}.

GIADA found compact particles ranging from 30\,\micron~to 1\,mm, and fluffy agglomerates ranging from 0.2\,mm to 2.5\,mm \citep{Fulle_2015}. GIADA also found that the dust size distribution changes as the comet orbits the sun; from a power law index of > -2 at 2.2 and 2.1\,au inbound, to -3.7 at perihelion \citep{Fulle_2016}. By comparing dust coma models to OSIRIS data, \cite{Marschall2020} found that the best fitting dust size distribution power law index at perihelion is around 3.7$^{0.57}_{-0.078}$ and that the smallest dust particles should be "strictly smaller than ~30\,\micron~and nominally even smaller than ~12\,\micron." \cite{Marschall2020}.

These values are mostly valid for the dust coma surrounding the comet. Retrieving similar values for the refractory phase before it leaves the surface is challenging. A derivation of a direct relationship between the coma dust size and surface dust size is difficult, as an unknown selection of mechanisms could take place in between (for example the sublimation of volatiles after ejection, leading to further fragmentation). Nevertheless, using images of the surface taken by the ROLIS instrument onboard Rosetta, \cite{Mottola2015} derived a dust size distribution power law index of -2.2 for agglomerates ranging from 4 to 0.2\,cm.

\subsubsection{Conventions for the definitions of particles}

For the definitions of the different types of particles we are analysing, we will use the convention defined in \cite{Guttler2019}. They define three categories of particles: 

\begin{enumerate}
    \item A "solid group", composed of irregular grains, monomers, and dense aggregates of grains. They are mainly characterised by a strong tensile strength and by a very low porosity (less than 10\%).
    \item A "porous group", composed of porous agglomerates of grains and clusters of porous agglomerates (porosity 10-95
    \item A "fluffy group", composed of fractal dendritic agglomerates, generally very porous (porosity $>95\%$) and with a very low tensile strength, see \citet{Fulle217} for more details. 
\end{enumerate}

For the CoPhyLab experiments, we want to match two of the particle categories to the known cometary values, the solid group and the porous group. In the solid group, we are unable to distinguish between an individual grain and a dense agglomerate of grains with the measurement methods at our disposal (see section below). We make the assumption, due to their low porosity and high strength, that the dense agglomerates are indistinguishable from the individual grains for most of our planned experiments. Thus, we hereafter use "grain" to mean individual grains and dense agglomerates of grains. In a similar way, we lack the ability to distinguish between porous agglomerates of grains and clusters of porous agglomerates. They will be both referred to hereafter as agglomerates, as the differences between them are not expected to impact any of the individual experiments. We choose not to try and match the fluffy group composed of fractal agglomerates, mainly because they are difficult to produce in the quantities needed for the experiments. Additionally it would complexify the comparison of numerical models to the experiments.

\subsubsection{Particle size measurement methods}
\label{sec:dustmethods}

Two methods were used to evaluate the grain size-frequency distribution of our samples. The first method involved using a sieving tower with levels of different-sized sieves. Using this method, the smallest size fraction reachable is 44\,\micron. The second method involved using a microscope and analysing the images of dust particles suspended in ethanol. The suspension was exposed to an ultrasound bath in order to separate the agglomerates into their individual grains. The images, taken at different zoom factors, were then analysed with a shape recognition script that identified individual grains and fitted a circle around them to provide an estimate of their size. In order to determine the smallest sizes visible by this method, we calibrated the microscope using a USAF 1951 Chart calibration target. Images of the target demonstrated that we could observe particles as small as 0.43\,\micron. It might be possible to observe smaller particles, but unfortunately we were not able to verify this as we already reached the limit of our calibration target. We therefore take 0.43\,\micron~as our best resolution using this method, but assume that we could resolve particles half that size.

\begin{figure}
	\includegraphics[width=\columnwidth]{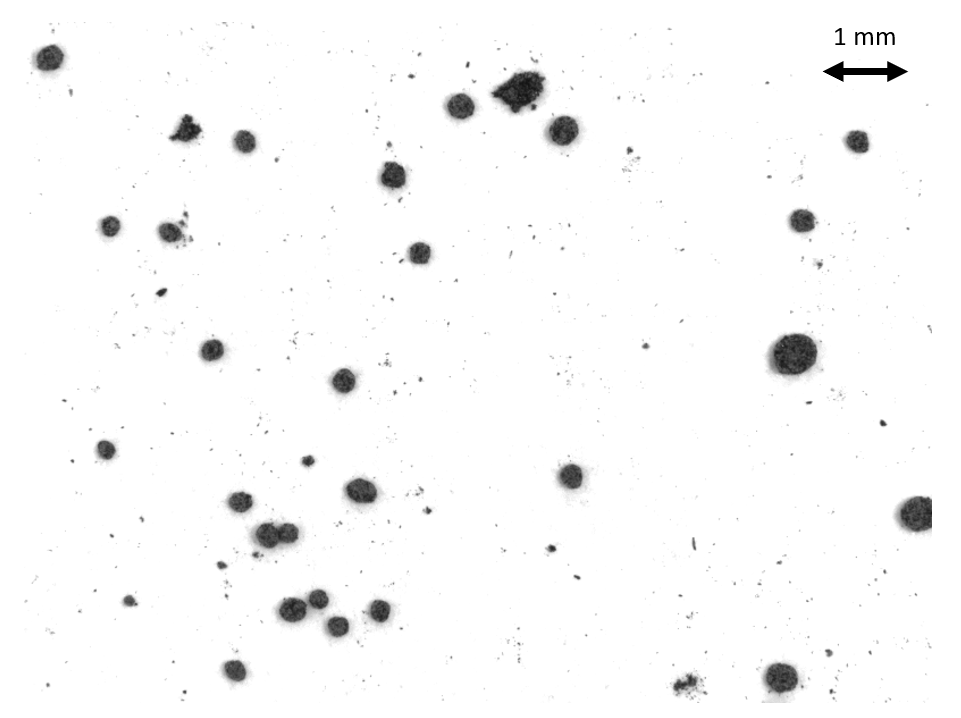}
    \caption{Example of one of the images of the dust taken with a camera and macro-lens used to derive the agglomerate size-frequency distribution}
    \label{fig:cameraImage}
\end{figure}

In order to measure the agglomerate size-frequency distribution (agglomerates that are formed from the shaking and combination of particles from both phases), we combine the microscope method (for agglomerates in the micrometre range) with a camera with a macro-lens (for agglomerates in the 100 micrometre range up to the millimetre range). The camera observes a white piece of paper on which we sprinkled the dust mixtures (see Figure~\ref{fig:cameraImage}). The images were analyzed with a shape detection algorithm to estimate their number and size, similarly to how the microscope images were analysed. Using this method the resolution is 24\,\micron.

\subsubsection{CoPhyLab dust grain size-frequency distribution}

The grain size-frequency distribution is known for both base materials. For pure silica, the dust size-frequency distribution was previously measured by \citet[][their Figure 3]{Kothe2013}, in which the mass-frequency distribution shows that most of the mass is contained in grains with sizes between 1\,\micron~and 10\,\micron. The grain size of the charcoal was measured in this work by using the microscope method described in the previous subsection. The particles were quite irregular, so we systematically took the longest dimension to be representative of the size of the particle.

\begin{figure}
	\includegraphics[width=\columnwidth]{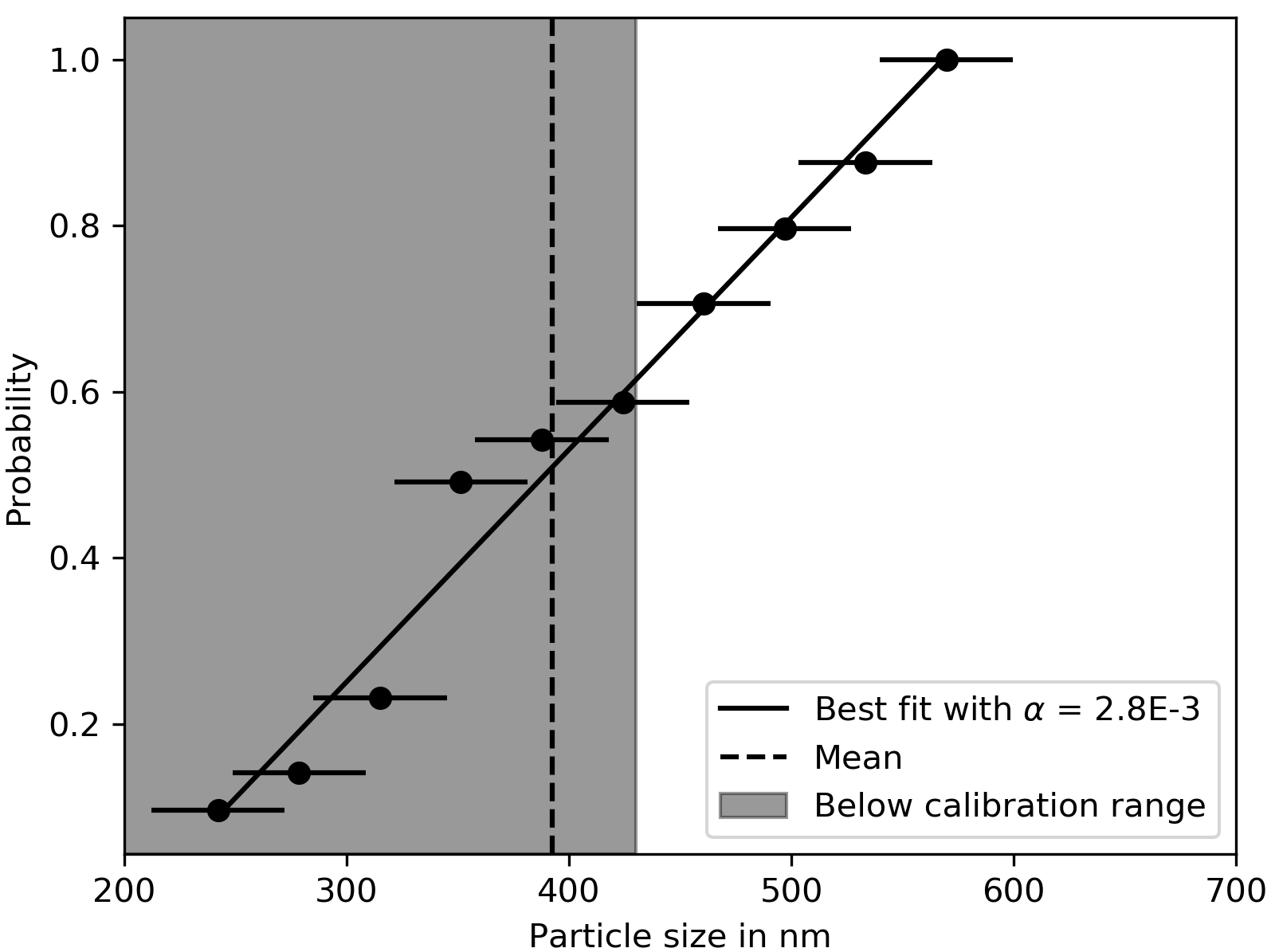}
    \caption{Cumulative size-frequency distribution of the grains (250\,nm < r < 600\,nm) of charcoal. The grain sizes are fitted with a linear function and the average size (392\,nm) is shown. We also indicated in the grey-shaded area the sizes that are below our minimum calibration size.}
    \label{fig:sizeDistribution}
\end{figure}

Figure~\ref{fig:sizeDistribution} displays the cumulative size-frequency distribution of the charcoal grains observed with the help of our microscope (250\,nm < r < 600\,\nm). This size-frequency distribution can be compared to Rosetta estimates for the grain size. For the charcoal grains we found a slope of the linear fit to our data of $\alpha = 2.8\,\times\,10^{-3}\,nm^{-1}$ (see Fig. \ref{fig:sizeDistribution}), close to those found by MIDAS ($\alpha = 3.5\,\times\,10^{-3}\,nm^{-1}$ for particle G and $\alpha = 3.2\,\times\,10^{-3}\,nm^{-1}$ for particle D \citep{Mannel2019} and VIRTIS ($2.5-3.0\,\times\,10^{-3}\,nm^{-1}$ \citealt{Bockel_e_Morvan_2017}).

The average grain size of the two components of the CophyLab dust are compatible with the smallest values detailed in the previous section and with the constraints given by \cite{Marschall2020} (grains with sizes less than 1\,\micron). Based on these results, we find that all possible mixtures of silica dust and charcoal have the required grain size.

\subsubsection{CoPhyLab dust agglomerate size distribution}

\begin{figure}
	\includegraphics[width=\columnwidth]{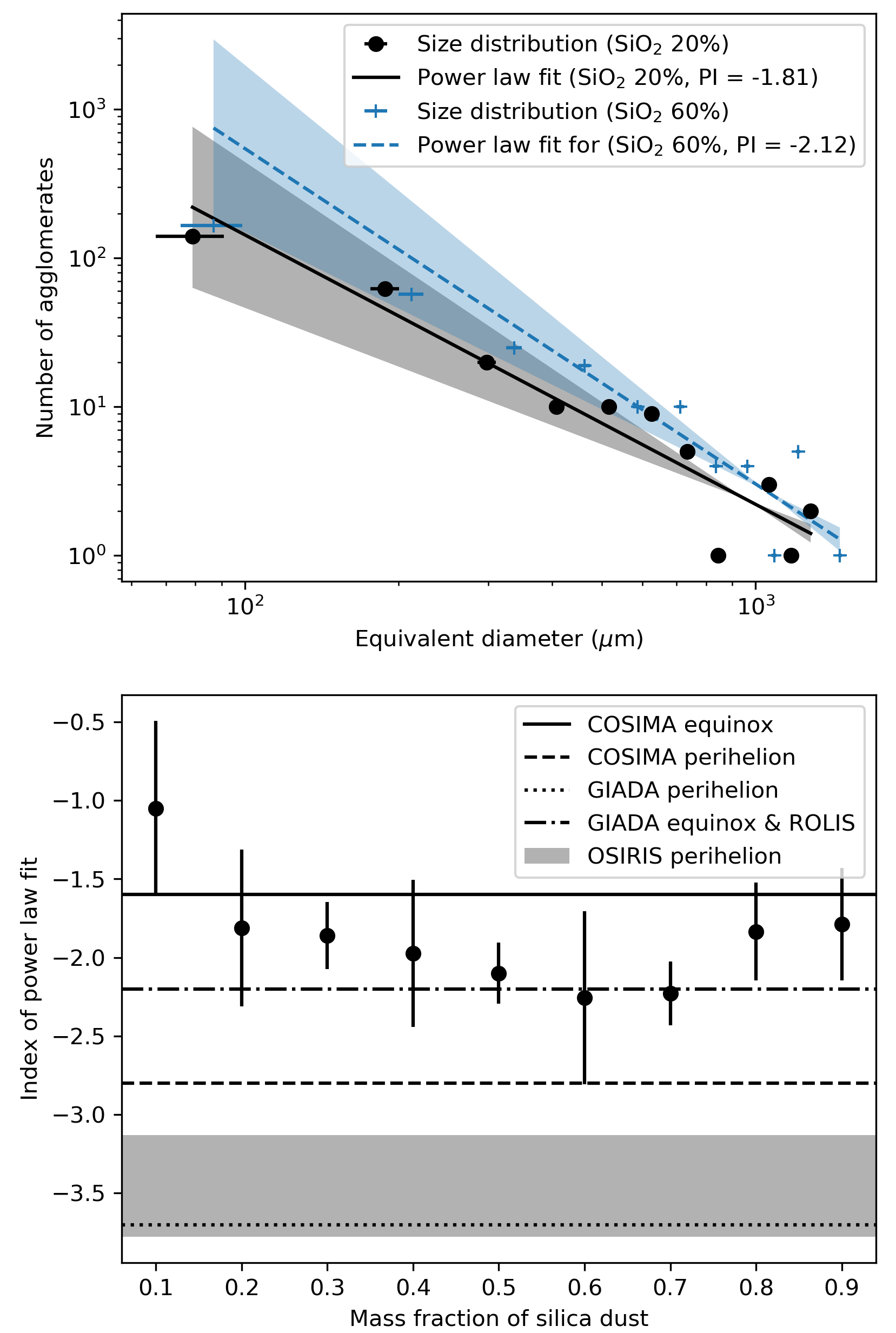}
    \caption{Top: Agglomerate size-frequency distribution between the 100\,\micron~and 2\,mm for two charcoal/silica mass ratios. The shaded areas represent the uncertainties of the fitted power law indices. Bottom: Index of the power law fit as a function of the mass fraction of silica dust. The error bars show the uncertainty of the fit parameters as shown in the top figure. The horizontal lines represent estimated values of the power law index derived from multiple Rosetta instruments described in Section~\ref{sec:cometarydustvalues}.}
    \label{fig:agglometratesizeDistribution}
\end{figure}

The CoPhyLab dust agglomerates were measured with the method described in Section~\ref{sec:dustmethods} for mass fractions of silica dust reaching from 10\,\% to 90\,\%. For each of the nine mixtures, the size-frequency distribution was fitted with a power law in the 100\,\micron~to 1\,mm size range (two examples can be seen in Figure~\ref{fig:agglometratesizeDistribution} top). In the bottom of Figure~\ref{fig:agglometratesizeDistribution}, the resulting power law indices are plotted as a function of the silica mass fraction and compared to values derived from measurements of the Rosetta instruments. We found that most mixtures (from 20\,\% to 90\,\% silica dust) have a power law index between -1.5 to -2.2, which is in good agreement with the dust size-frequency distribution found by GIADA, ROLIS, and COSIMA at equinox. However, we do not reach the indices determined by the instruments at perihelion. This implies that to compare our results to the most active phases of the cometary orbit, different size distributions are probably required. 

However, this comparison has a few caveats. The COSIMA power law index was determined mostly for particles with sizes between 30\,\micron~to 150\,\micron, which is a an order of magnitude lower than our agglomerate sizes \cite{Merouane2017}. Another possibility is that the agglomerates broke upon impact and only the resulting fragments are observed. Additionally, all of these power law indices are derived from dust that has left the cometary surface and thus does not necessarily represent the properties of the dust on the nucleus. For these reasons, we are more confident that the ROLIS power law index is the most representative of the nucleus cometary dust. It is interesting to note that GIADA equinox observations also find a similar power law index to that of ROLIS \citep{Fulle_2016}. 

Taking all this information into account, we find that most of our mixtures possess power law indices very close to those found by GIADA and ROLIS. In four cases (20, 40, 60, and 70\,\% of silica dust), the error bars overlap with the hypothesized surface dust power law index. In three cases (30, 80, and 90\,\% of silica dust), the values are found to be sufficiently close and thus are considered acceptable mixtures. The only outlier is the 10\,\% silica dust mixture. Therefore, we decided to not consider it further for our CoPhyLab experiments.

\subsection{Material density}
\subsubsection{Cometary dust densities}
\label{sec:knowndensities}

The material density of cometary dust is a fairly unknown quantity. The Rosetta Radio Science Investigation (RSI) \citep{Patzold2016} measured that the comet as a whole has a bulk density of 533\mmp6\dst, which implies a high porosity of 72–74\,\%. In order to derive this porosity value, \citet{Patzold2016} used a value of 2600\mmp400\dst as the material density for the cometary dust (density with no porosity). \cite{JORDA2016}, using an estimated global density of the comet of 532\mmp7\dst and a grain density between 2000 to 3500\dst, found that the comet has a porosity between 70\,\% and 75\,\%. For their analysis of the data from the Comet Nucleus Sounding Experiment by Radiowave Transmission (CONSERT), \cite{Kofman2015} used material densities of 2500 to 3500\dst (based on densities of ordinary and carbonaceous chondrites) and found that the nucleus has a porosity of 75 to 85\%. In a later study, the CONSERT team extended this range from 2000 to 3500\dst \citep{Herique2017}. In order to be compatible with their measurements, the authors hypothesized that a high fraction of carbonaceous material is needed. This in turn led them to favour materials with densities closer to 2000 than to 3500\dst (see Table 7 in \citealt{Herique2017} with the compatible materials in bold). They find that CR2 chondrites could be discarded due to the lack of hydrated minerals in 67P/C-G.

\subsubsection{Density measurement method}

In order to measure the intrinsic material density (or grain density) of our analogue materials, we used a helium pycnometer (Upyc-1200e-V5.04, located at the University of Bern) that measures the volume of helium needed to fill the voids in a sample of known volume. This then allows us to calculate a material density for the sample. This method has been previously used to measure the material density of Lunar and Martian simulant regolith samples \citep{Mattei2014, Brouet2019, Pommerol2019}.

\subsubsection{CoPhyLab dust density}

\begin{figure}
	\includegraphics[width=\columnwidth]{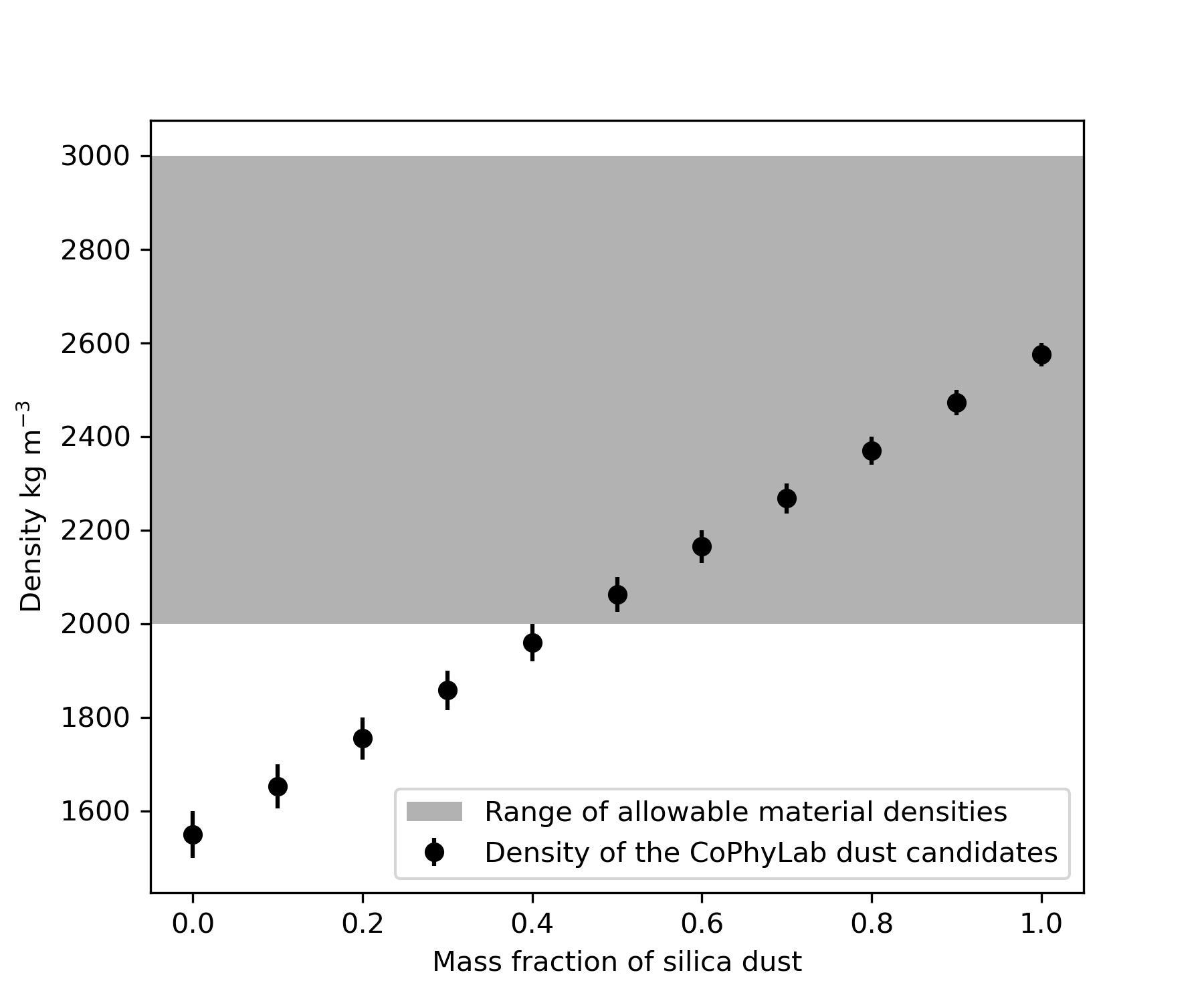}
    \caption{Density of the CoPhyLab silica-charcoal dust mixtures as a function of the mass fraction of the silica dust. The grey-shaded area indicates the allowable range for the material densities of a cometary dust analogue \citet{Kofman2015,JORDA2016}}
    \label{fig:density}
\end{figure}

Using the method mentioned above, the material densities of the silica dust and the charcoal end members were determined to be 2550-2600\dst and 1500-1600\dst, respectively. For the analogue mixtures, we used a linear mixing law in order to determine the material densities. These densities are plotted in Figure~\ref{fig:density}. In this Figure, we also indicated the range of allowable material densities for our mixtures (those used by previous studies as estimates for the 67P/C-G refractory grain density). In order to take into account as many of the densities described in Section~\ref{sec:knowndensities}, we choose a range from 2000 to 3000\dst. We can see from Figure~\ref{fig:density} that all mixtures with less than 50\,\% silica dust by mass do not have the minimum density value required. We also note that our densities are generally quite low (< 2600\dst), making them compatible with the values in \cite{Herique2017}.

\subsection{Gas permeability}

 Gas permeability is an important factor in numerical models for the activity of comets as it describes how easily solid material allows gas (like sublimated water molecules) to travel through the pores of the surface layers (see \citealt{Gunlach2020b} for an example). The gas flow through a porous material can be represented by two main parameters. The gas permeability that represents the viscous flow of gas and the Knudsen diffusion coefficient.

\subsubsection{Cometary dust gas permeability}

The Philae lander was designed to measure various surface properties with several sensors. Unfortunately, none of those were dedicated to the measurement of gas permeability. Part of the CoPhyLab experiments are devoted to better constrain this property for cometary physical analogues. Consequently, the gas flow measured for the cometary-analogue candidates is presented here but is not used to constrain the best analogue of all the candidates.

\subsubsection{CoPhyLab dust gas permeability measurement method}
To measure gas permeability, we performed gas flow experiments using a cylindrical sample container filled with CoPhyLab dust and mounted in the middle of a vacuum chamber. The sample container separates the chamber into two individual compartments (upstream and downstream compartment), connected only via the sample. With this chamber configuration, the gas streaming into the upstream compartment is directed through the sample into the downstream compartment, which is evacuated by pumping. The gas used for the measurements is cleaned oil-free air. The gas flow is controlled by a thermal mass flow sensor, allowing for gas flow in the range of 0.15\mgs to 20\mgs. This setup was previously used to measure the gas permeability for cometary analogues \citep{Schweighart2021}.

Three different types of pressure sensors were used to achieve high-accuracy measurements within different pressure ranges, one of each type connected to the upstream and downstream compartment. A sieve at the bottom of the cylindrical sample container supports the weight of the sample, holds it in place and ensures that gas can pass from one compartment to the other. To measure the gas permeability, a constant flow of gas is directed through a sample of known volume and the pressure in the two compartments is measured as a function of the gas flow. During the evacuation process, the difference in pressure between the upper and lower compartment is kept as low as possible to minimise mechanical changes in the sample (although the building up of compacted layers at the bottom of the sample could still not be avoided). After a minimum pressure of 10$^{-2}$\,mbar was obtained, the chamber was flooded with gas, starting with a gas flow of 0.15\mgs. As the system adapted to the change, the pressure approached a steady state. Whenever a steady pressure level was reached, the mass flow was doubled (more details can be found in \citealt{Schweighart2021}).

Although the influence of parts other than the sample are minimised, there is still some influence from the sieve at the bottom of the container. Pressure differences for an empty sample container with just the sieve were measured and the pressure values obtained during the measurement with different samples were corrected to account for the pressure differences caused only by the sample. All measurements were done at room temperature.

Samples were prepared as described in Section~\ref{mixingrecipie}. For the different samples, the sizes of the individual dust agglomerates were in the range of 56\,\micron\, to 1000\,\micron. 

Gas permeability and Knudsen diffusion coefficient were calculated from the measured pressure values (see \cite{Schweighart2021} for further details).

\subsubsection{CoPhyLab dust gas permeability}

As mentioned above, there are no in-situ measurements of the gas flow through cometary surface dust layers available. To get a first impression of possible gas activities, we performed gas flow measurements with one of the SiO$_2$/charcoal mixtures (70\,\%\,/\,30\,\%). The samples are difficult to handle and, as will be seen later, the results show a strong dispersion. Samples with a height >10\,mm tend to build up a compacted layer at the outlet (downstream side of the sample) and often channels are created close to the sidewalls of the sample container, resulting in pits on the surface. Samples with a higher amount of charcoal have cracks that build up in the middle of the sample. Such measurements were excluded from further data evaluation.

Some of the changes in the samples take place during the evacuation process before the gas flow is started. To minimise mechanical changes of the sample, the pressure difference between the upper and the lower compartment should be kept low, preferably at less than 1\,mbar. This is hard to achieve with the current set-up, therefore, the number of samples whose measured properties can be used for data interpretation is limited. In the same way as the experiments done with glass beads by \citealt{Schweighart2021}, one can see a dependency of the gas flow on the size of the single agglomerates. Although the average size-frequency distribution of samples is known (see Section~\ref{sec:cometarydustvalues}), small variations can lead to non negligible changes in the gas permeability measurements.

For the CoPhyLab dust sample with $70\,\% SiO_{2}$ and $30\,\%$ charcoal (sample height $8\,$mm, porosity = $75$\mmp 1.5\,\%, maximum agglomerate size in the range $700\,\micron - 900\,\micron$) we find an average gas permeability of $B = 2.04\cdot$\,$10^{-11}$\,$m^2$\mmp$ 0.22\cdot$\,$10^{-11}$\,$m^2$ and a Knudsen diffusion coefficient $D_{K}$ of $1.74\cdot$\,$10^{-3}$\,$m^2$\,$s^{-1}$\mmp $0.24\cdot$\,$10^{-3}$\,$m^2$\,$s^{-1}$. 

\subsection{Spectral properties}

\subsubsection{Cometary dust spectral properties}

The spectral properties of planetary bodies can help us to identify the composition and physical properties of their uppermost surface layers. Studying the spectral properties of cometary analogues, especially when mixed with water ice and exposed to solar illumination, can help us to better understand processes which take place on a comet. To this end, we plan to study the spectral properties of our CoPhyLab dust samples exposed to simulated cometary conditions (temperature, pressure and illumination).

The spectral reflectance of a closely-packed particulate medium is a function of the viewing geometry, the optical properties of the scatterers that constitute the medium and the physical properties of the medium itself (e.g. the surface porosity and the surface roughness). Hence, for this study, we aimed at characterizing the overall spectral properties of the pure samples and mixtures.

The spectral properties of comets have been measured at multiple occasions. For comparison to our analogues, we mainly focus on the spectra acquired by VIRTIS of the Rosetta spacecraft as they are the most recent and detailed spectral properties we have to this day. These studies have revealed the presence of water ice among a complex mixture of organics and refractory materials on the nucleus' surface \citep{Capaccioni_2015, Quirico_2016, Poch_2020}. These VIRTIS measurements also allowed the detection of overnight surface deposition of volatiles released during the local daytime and in particular to characterise the diurnal cycle for water-ice \citep{Sanctis_2015}.  

\subsubsection{Spectral properties measurement method}

The spectral properties of the samples were investigated using the University of Bern's Mobile Hyperspectral Imaging System (MoHIS \citealt{Pommerol2019}). This instrument uses a 250\,W halogen lamp as a light source, to which a monochromator (Oriel, MS257) is attached, with three gratings to sweep from the NUV (Near UV) to VNIR (Very Near InfraRed) domains and select specific wavelengths with a narrow bandpass. An optical fibre is placed after the monochromator and set up to illuminate the sample perpendicularly to its surface. A CCD camera is used for the observations in the visible domain (380-940\,nm) and an MCT (Mercury, Cadmium, Telluride) detector is used for the infrared observations (950-2450\,nm). Additional details on the setup can be found in \citet{2019Kreuzig}.

\subsubsection{CoPhyLab dust spectral properties}

\begin{figure}
	\includegraphics[width=\columnwidth]{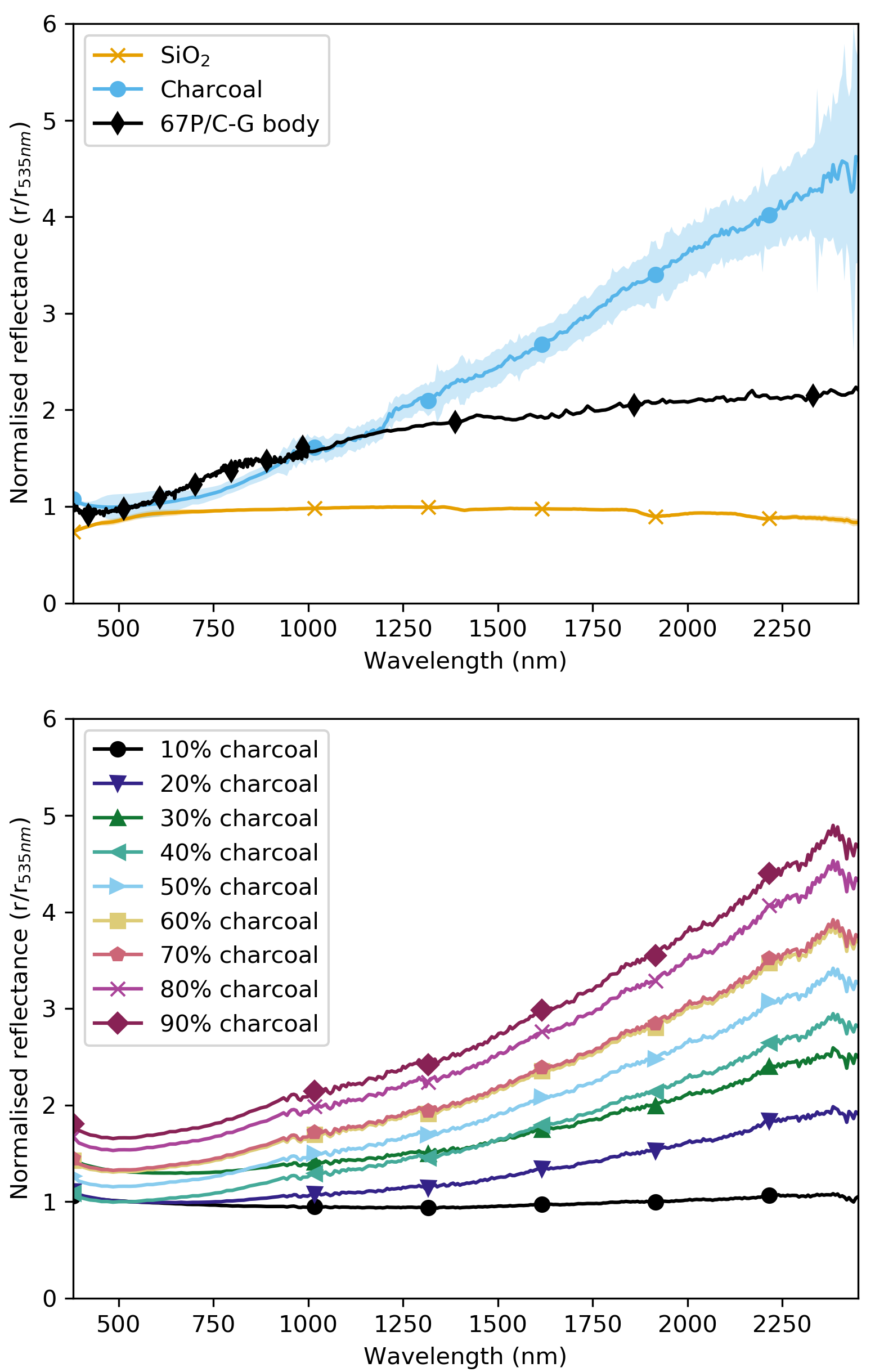}
    \caption{Top: Reflectance spectra of the pure phases of the Cophylab dust with normalisation at 535\,nm. The shaded area represents the dispersion associated with these measurements (present for both SiO$_2$ and charcoal but the SiO$_2$ normalised error is small). The 67P/C-G spectrum corresponds to a spot of the Aswan terrace on comet 67P/C-G \citep{Capaccioni_2015}. Bottom: Reflectance spectra of Cophylab dust mixture with normalisation at 535\,nm. Symbols included for readability.}
    \label{fig:em_spectra}
\end{figure}

Nine mixtures with a 10\,\% increment by mass of charcoal were prepared. The sample holders were filled with the mixture using a standard kitchen sieve (400-500\,\micron~mesh) to achieve a smooth and featureless uppermost layer. The spectral properties in the visible and near-infrared domains (from 380 to 2450\,nm) are presented in Figure~\ref{fig:em_spectra}.

These measurements show an important difference between the SiO$_2$ and the charcoal powder in the visible and infrared domain (see Figure~\ref{fig:em_spectra}). Across the 400-2500\,nm wavelength range, the reflectance factor of the SiO$_2$ sample has a median value of 0.96 and reaches close to unity near 1250\,nm. SiO$_2$ displays a noticeable positive slope in the visible range and shows an overall $\sim$ 15\,\% decrease in reflectivity across the 1.2-2.5\,\micron wavelength range. We note the presence of hydration and hydroxylation features around 1.4\,\micron~($\sim$ 7140\,cm$^{-1}$), 1.9\,\micron~($\sim$ 5250\,cm$^{-1}$), as well as around 2.2\,\micron~($\sim$ 4545\,cm$^{-1}$) \citep[see][and references therein]{Elliott_1971, Lipp_1992, Pommerol_2008}. It is to be noted that the samples were measured at ambient conditions. A thorough heating of the SiO$_{2}$ material, and its preservation in a desiccating oven prior to manipulations, could certainly lead to a decrease of the intensity of these features. Overall, the spectrum of pure SiO$_{2}$ powder does not present any large variations and remains at most within 15\,\% of its reflectivity value at 535\,nm \footnote{wavelength of the central wavelength of the green filter of the Rosetta/OSIRIS Narrow Angle Camera.}. We also displayed in this figure the spectrum of comet 67P/C-G (specifically the spectrum of the Aswan terrace on the nucleus \citep{Capaccioni_2015} and any SiO$_2$ relative spectral variations appear very moderate when compared to it.

The spectrum of the charcoal is characterised by its lack of features and its low reflectivity, which remains below 5\,\% in the visible domain and increases with a slope of 18\mmp 9\,\%/100\,nm between 800 and 2400\,nm, where it reaches a reflectivity of 13\mmp 5\,\%. The normalisation of the spectrum at 535\,nm highlights this increase and also points to a slight reflectivity decrease in the 400-500\,nm region (with a spectral slope of -4\mmp 1\,\%/100\,nm), as well as the presence of a broad but shallow dip in the charcoal's spectrum between 1\,\micron and 1.2\,\micron. This reflectance dip is compatible with the absorption band of CH bonds present in organic compounds \cite{Izawa_2014}. While the spectrum of charcoal and 67P/C-G are similar in the 400-550\,nm and the 950-1200\,nm ranges, they differ across the 600-900\,nm domain and throughout the investigated infrared domain.

In examining the spectra of the mixtures (Figure~\ref{fig:em_spectra}, bottom), we observe that the charcoal powder drives their spectral properties. This is illustrated by the low reflectance of all mixtures, including those that have small fractional contents of charcoal powder, as well as by the overall absence of spectral features and the monotonically positive average spectral slope in the near-infrared. The reflectance is close to 72\,\% for the 10\,\% charcoal mixture and close to 88\,\% for the 20\,\% charcoal mixture. Comparing the spectra of the mixtures with that of 67P/C-G, normalised at 535\,nm, we are able to again observe the same differences between the charcoal powder and 67P/C-G.

In the visible domain, the relative reflectance values of the mixtures are comparable to that of 67P/C-G mostly around the 500-570\,nm window, with an overall mismatch in the near-infrared domain. However, all mixtures with more than 20\,\% charcoal content appear to match the relative reflectance of 67P/C-G at certain wavelengths: around 2\mc for the 30\,\% charcoal mixture, around 1.78\mc for the 50\,\% mixture, around 1.5\mc  for the 70\,\% mixture, and around 1.35\mc for the 90\,\% charcoal mixture.

For each sample, we computed spectral slopes between 535\,nm and 882\,nm, using the formula from \cite{Delsanti_2001}. The results are reported in Table~\ref{tbl:spcslp}. In general, the increase of the spectral slopes appears to follow the increase in charcoal fraction. The difference between the spectral slope of the 80\,\% charcoal mixture and its neighbours could be potentially explained by a spatial inhomogeneity within the mixture, further enhanced by the low reflectance of the mixture and the reflectance dispersion at each wavelength. This would also explain its similarity to the spectral slope of the pure charcoal. The spectral slopes of the 10\,\% and 20\,\% charcoal mixtures are lower than even that of the pure SiO$_2$ sample. Moreover, while the reflectances of these mixtures transition down to the reflectance level of the pure charcoal sample, their spectral slopes appear to be negative or neutral over the visible range (Figure~\ref{fig:em_spectra}). However, they exhibit markedly positive spectral slopes beyond 1.2\,\micron, similarly to the pure charcoal sample (Figure~\ref{fig:em_spectra}). The lower-than-average spectral slopes of the 10\,\% and 20\,\% charcoal mixtures in the visible could be explained by the scattering contribution of the SiO$_2$, which is then still the main species of the mixture.

\begin{table}
  \centering
  \caption{\label{tbl:spcslp} Visible spectral slope of the pure CoPhyLab dust materials and their mixtures using the formula from \citet{Fornasier_2015}, and BVRI (Blue, Visual, Red and Infrared) colours computed using Bessell filters.}
  \begin{tabular}{|c|c|}
    \hline
    Sample & 535--882 nm slope \\
    $ $    & (\%/100 nm)\\
    \hline
    0\% (Pure SiO2)  &  2.8\mmp0.2 \\
    10.0\,\%   & -1.4\mmp0.2 \\
    20.0\,\%   &  1.0\mmp0.3 \\
    30.0\,\%   &  3.8\mmp0.9 \\
    40.0\,\%   &  5.0\mmp1.0 \\
    50.0\,\%   &  4.0\mmp1.0 \\
    60.0\,\%   &  5.0\mmp1.0 \\
    70.0\,\%   &  5.0\mmp1.0 \\
    80.0\,\%   & 11.0\mmp3.0 \\
    90.0\,\%   &  7.0\mmp2.0 \\
    100\% (Pure charcoal) & 10.0\mmp3.0 \\
    \hline
  \end{tabular}
\end{table}

Compared to the nucleus of 67P/C-G, these computed spectral slopes are lower than the average terrain of the comet. In the high-spatial observations of the nucleus near opposition \citep{Feller_2016, Feller_2019}, the spectral slope of the typical boulder-less terrain was close to 18\,\%/100\,nm, i.e. about 2.6 times greater than the spectral slope of the 90\,\% charcoal mixture. On the nucleus, areas with low spectral slopes were also observed to be brighter than their surroundings, and were typically associated with material fractionally enriched in water-ice \citep{Barucci_2016, Hasselmann_2017, Deshapriya_2018}.

\subsection{Photometric properties}

\subsubsection{Cometary dust photometric properties}

The photometric properties of the considered end-members and mixtures were investigated and are presented in the companion paper Feller \textit{et al.} (in preparation). We focus here on the single-scattering (w$_\mathrm{ssa}$), the geometrical (A$_\mathrm{geo}$) and on the bidirectional albedo (A$_\mathrm{bd}$). These are important quantities to the study of the physics of our samples as they describe how dark our samples are and how much total energy is absorbed through illumination. The single-scattering albedo corresponds to the propensity of grains/aggregates to scatter light through simple reflection. The geometric albedo at incidence and emergence zero corresponds to how much light is scattered at opposition. Finally, the bidirectional albedo is the bond albedo before integration over all wavelengths, it represents the reflectance integrated over the incidence hemisphere and the emergence hemisphere.

In Table~\ref{tab:bond_cmp}, we compare the albedos of selected CoPhyLab dust mixtures (see below) with those of asteroids and comets, determined using the Hapke reflectance model. \citet{Li_2007,Li_2013} modelled the photometric properties of comet 9P/Tempel~1 as observed in the Bessell V-Band filter \cite{Bessell_1990} by the Deep Impact and Stardust-NExT cometary missions respectively. \citet{Helfenstein_1989} fitted the phase curves of a collection C-type asteroids as observed from Earth in the V-Band filter. More recently, the Rosetta mission allowed us to study the nucleus of comet 67P/C-G at great detail, in particular \citet{Fornasier_2015} used low- and large-phase angle observations acquired with a V-band equivalent broadband filter during the orbiting with the nucleus to determine its 
photometric properties.

 \begin{table}
  \caption{\label{tab:bond_cmp} Single-scattering (w$_\mathrm{ssa}$), geometric (A$_\mathrm{geo}$), and bidirectional (A$_\mathrm{bd}$) albedos at, or close to, 550\,nm for selected CoPhyLab dust mixtures compared to values of comets and asteroids.}
  \begin{small}
  \begin{center}
  \begin{tabular}{|c|c|c|c|}
   \hline
    $ $     & w$_{ssa}$ (\%) & A$_{geo}$ (\%) & A$_{bd}$ (\%)\\
   \hline
   10\,\% charcoal        &47.8\mmp0.3 &  34\mmp2   & 22.7\mmp0.1 \\
   30\,\% charcoal        &11.7\mmp0.1 &   9\mmp1   &  5.3\mmp0.1 \\
   50\,\% charcoal        & 9.0\mmp0.1 &   6\mmp1   &  3.7\mmp0.1 \\
   70\,\% charcoal        & 7.6\mmp0.1 &   5\mmp1   &  3.2\mmp0.1 \\
   90\,\% charcoal        & 5.7\mmp0.2 &   4\mmp1   &  2.5\mmp0.1 \\
   \hline
   9P/Tempel 1$^{a,b}$  & 3.9\mmp0.5 & 5.6        & 1.3          \\
   67P/C-G$^{c}$       & 3.7\mmp0.2 & 5.9\mmp0.3 & 1.23\mmp0.01 \\
   \hline
   Average C-type$^{d}$ & 2.5        & 4.9        & 0.8*  \\
   \hline
  \end{tabular}
  {\newline \small References - a:\citealt{Li_2007a}; b: \citealt{Li_2013}; 
    c: \citealt{Fornasier_2015};\\
    d: \citealt{Helfenstein_1989}}\\
    *: anisotropic case estimation, using Eq. 11.21 of \citealt{Hapke_1993}
  \end{center}

  \end{small}
 \end{table}

\subsubsection{Photometric properties measurement method}

The photometric phase curves of our samples were measured using the PHysikalisches Institut Radiometric Experiment 2 (PHIRE-2) radiogoniometer. We refer the reader to \citet{Pommerol_2011} for a complete description of the instrument and the calibration process. The measurements were acquired in the principal plane (i.e. the azimuth angle equals either 0$^{\circ}$ or 180$^{\circ}$) using the instrument's 550\mmp\,10\,nm broadband filter. The light source was set at an incidence angle of 0$^{\circ}$ and measurements were acquired every 5$^{\circ}$ in the plane of emergence from 5$^{\circ}$ to 75$^{\circ}$. The phase curves of the samples cover thus the 5$^{\circ}$--75$^{\circ}$ phase angle range.

\subsubsection{CoPhyLab dust photometric properties}

Fitting an implementation of the so-called Hapke photometric model \cite{Hapke_1993} to the obtained phase curves (see the companion paper Feller \textit{et al.} (in preparation), we were able to derive the single-scattering, geometrical and bidirectional albedo values of the end-members and the CoPhyLab dust mixtures. As noted previously with their respective spectra, the single-scattering albedos highlight the respective brightness and dimness of the two pure materials (w$_\mathrm{ssa, SiO_2}~\simeq~$97.5\mmp0.1 and w$_\mathrm{ssa,charcoal}~\simeq~$5.6\mmp0.3). For comparison, the SiO$_2$ value is almost as large as that of ice \citep{Warren_2008,Yoldi_2015}, while the juniper charcoal's w$_\mathrm{ssa}$ is almost four times lower than that of lunar Maria \citep{Sato_2014}.

Generally the single-scattering albedo values decrease sharply with the increase of charcoal content, such that adding 10\,\% by mass of charcoal to a pure SiO$_2$ sample slashes its bidirectional albedo (83.4\mmp0.1) while adding 30\,\% by mass of charcoal reduces it by about 94\,\%. This sharp brightness decrease, also noted for the mixtures' spectra, is consistent with the findings by \citet{Warren_1980, Pommerol_2008} and \citet{Jost_2017a}, 
which respectively considered mixtures of snow (or fine-grained water ice) and soot (or ground anthracite) and pointed to the driving influence of the fine opaque particles over the photometric properties of the mixtures, even when amounting to a few percent of the total mass. This trend, which clearly stands out for bidirectional albedos, holds as well for the single-scattering and geometric albedo values of these dust mixtures.

Based on the determined geometrical albedos, while prepared dust mixtures with a charcoal fraction lesser than 50\,\% would, in comparison, appear brighter than the nucleus surface of 9P/Tempel 1 or 67P/C-G, those with a higher fraction of charcoal would appear at opposition as dark as their nuclei's surfaces. The further consideration of the bidirectional albedos points out that the investigated mixtures are inherently slightly better scatterers than the surfaces of these comets under all viewing geometries, as emphasised by the larger single-scattering albedo values of these mixtures. This apparent discrepancy between the geometrical and bidirectional albedo values of the dust mixtures and their cometary counterparts can be explained through a more extended consideration of the photometric behaviour of these surfaces (see Feller \textit{et al.}, in preparation). In essence, the comparable geometrical albedo value of the 67P/C-G surface with respect, for instance, to the 70\,\% charcoal dust mixture, is due to the strong backward-scattering nature of the nucleus surface reinforced at small phase angles the marked disappearance of shadows cast by dust grains and aggregates on the nucleus surface itself (i.e. the shadow-hiding contribution of the opposition effect, 
\citealt{Fornasier_2015, Feller_2016}). By comparison to other small bodies of the Solar System, at this wavelength, the prepared mixtures would be brighter than the average primitive C-type asteroid.

While the non-linear reflectance surge of the opposition effect is positively correlated with a surface roughness and particle grain size (\citealt{Deau_2013} and references therein), the mixture's scattering asymmetry is a function of the material driving the mixture's photometric properties \citep{Pilorget_2015} and therefore not only dependant on its physical properties, but also on its chemical composition, which in the case of 67P/C-G is still the subject of research \citep{Quirico_2016, Poch_2020}. Alternatively, in order to lower the dust mixtures' albedos to better match the values of cometary nuclei, more absorbent material, such as carbon black, can be used. Unfortunately, such a material is difficult to handle safely, due to its small particle size, and would present a potential danger for little gain. We therefore argue that the low albedos reached by mixture with a carbon mass content higher than 30\,\% are close enough (less than 5\,\% difference) to cometary values to allow us to model correctly the corresponding physics knowing the intensity of the illumination input.

\subsection{Mechanical properties}
\label{sec:knowntensilestrengths}
Characterising the mechanical properties of cometary material is important to understand the activity of comets, planetary formation, and the evolution of a comet's morphology. For example, dust ejection is only possible when the gas pressure created by the sublimation of volatiles in the nucleus of comets exceeds the tensile strength of the material \citep{Skorov_2012,Fulle2020, Gundlach2020}. These mechanical properties can also control the collapse of cliffs observed on the surface of the nucleus \citep{Groussin2015,Prialnik2017,Attree2018,Kappel2020}. It is important to distinguish between two possible tensile strengths. First, the inner-agglomerate tensile strength (referred to hereafter as inner tensile strength), which represents the strength of the binding of the solid particles inside the pebbles, and second, the inter-agglomerate tensile strength (referred to hereafter as inter tensile strength), which represents how strongly the agglomerates are bound to each other.

\subsubsection{Cometary dust mechanical properties}

With the arrival of Rosetta at comet 67P/C-G, knowledge on the mechanical properties of cometary matter has greatly increased. Two instruments located on the Philae lander were dedicated to the study of these properties on the surface close to the lander. First, the MUPUS penetrometer was not able to penetrate the surface at Abydos and \citet{Spohn2015} concluded that the uniaxial compressive strength of the material had to be at least 2\,MPa in order to prevent the penetration of MUPUS. Second, the Acoustic Surface Sounding Experiment (CASSE) was used to determine the mechanical properties at the first landing location, Agilkia. Based on its measurements, a compressive strength between 3.5 and 12\,kPa was determined \citep{MOHLMANN2018}. CASSE was also used to listen to the MUPUS penetrator as it attempted to penetrate the surface. This allowed estimates of the sheer modulus between 3.6\,MPa and 346\,Mpa, and of the Young's modulus between 7.2\,MPa and 980\,MPa \citep{KNAPMEYER2018}.

Another method used to determine the mechanical properties was based on geomorphological observations by the OSIRIS cameras to infer appropriate values based on analytical or numerical models. Geomorphological models of boulders and cliffs on the surface resulted in estimates of the inter tensile strength on the order 1-10\,Pa \citep{Kappel2020}. Using OSIRIS observations and examining the relationship between gravitational slopes and surface morphology, \citet{Groussin2015} estimated the compressive strength to be between 30 and 150\,Pa and the inter tensile strength to be between 3 and 15\,Pa. \citet{Vincent2017} performed a statistical analysis of large geomorphological features on the surface to calculate values of the inter tensile strength of 1-2\,Pa at a ten-metre scale. Comparatively, analyses of the morphology of overhanging cliffs on the nucleus constrained the range of possible inter tensile strengths to between 0.02-1.02\,Pa \citep{Attree2018}. In a recent study of the multiple landings of Philae, \citet{Laurence2020} found that the compressive strength of the surface material was lower than 12\,Pa. Theoretical values for the inter tensile strengths were derived by \citet{Skorov_2012} in their Figure 4.

\subsubsection{Mechanical properties measurement method}

A complete study of the mechanical properties of the CoPhyLab dust analogues is outside of the scope of this paper, as this will be performed through a dedicated experiment in a future study \citep[see][for details]{BLUM2014b}. At the time of writing, only the inner tensile strength of the CoPhyLab samples can be measured with the setup available. The method used by CoPhyLab is the Brazilian Disk Test (BDT) in which a cylindrical sample of material is formed and compressed on a scale with a thin metal loading bar. The force exerted on the sample is measured until the cylinder fractures. The inner tensile strength $\sigma$ can then be derived by

\begin{equation}
\sigma = \frac{2F}{\pi dl},
\end{equation}

$F$ is the maximum exerted force at break-up, $l$ is the thickness of the dust cylinder, and $d$ is the diameter of the sample, respectively. This method has been previously used to study protoplanetary and cometary analogues (see \citealt{Haack2020,Bischoff2020,Sebastian2020}).

\subsubsection{CoPhyLab dust tensile strength}

\begin{figure}
	\includegraphics[width=\columnwidth]{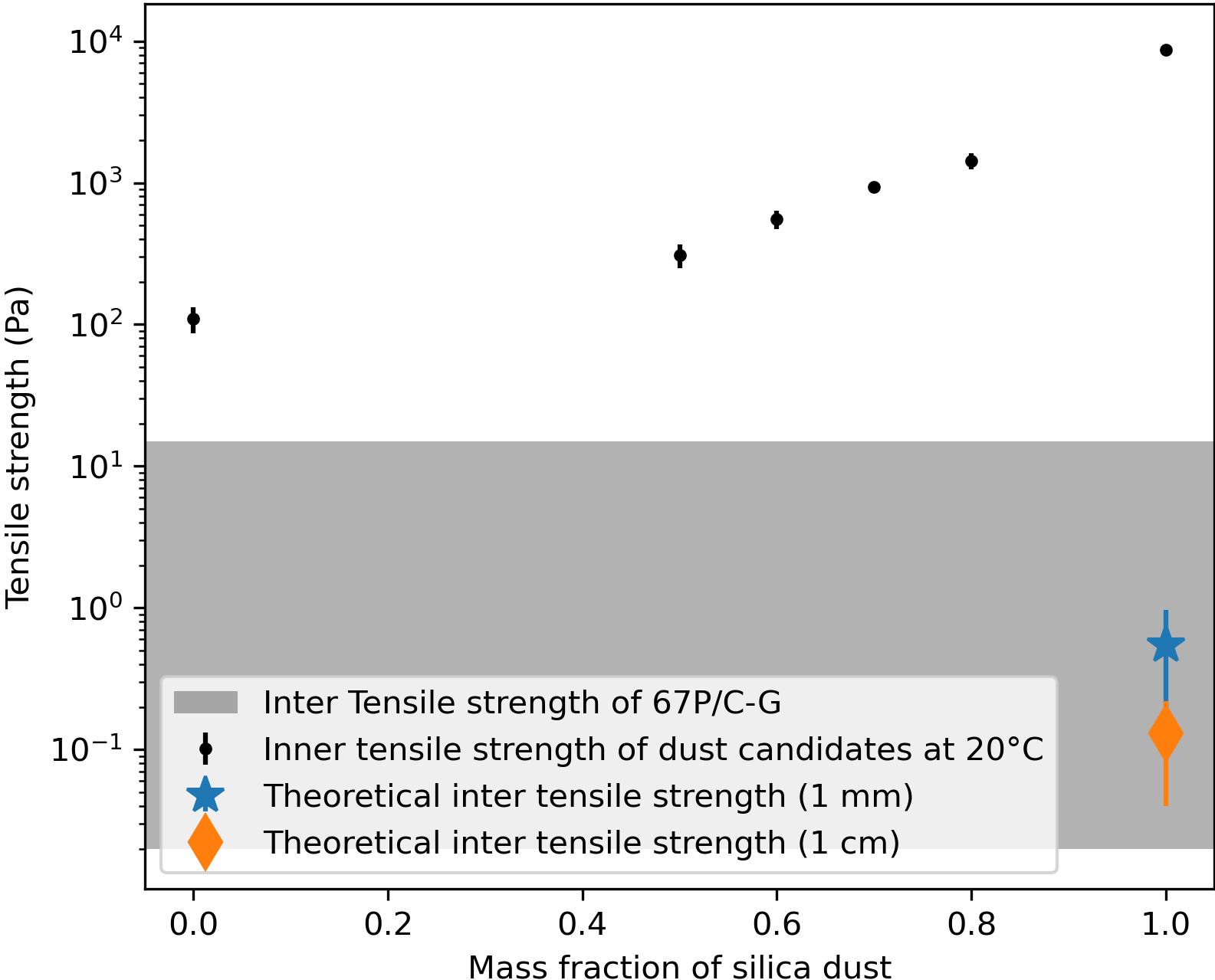}
    \caption{Tensile strengths of the CoPhyLab dust mixtures as a function of the mass fraction of silica dust with error bars (for some mixtures these are very small and therefore barely visible). We also indicate the allowable range for the inter-agglomerate tensile strengths of a cometary dust analogue in the grey-shaded area. The two theoretical cometary dust aggregates tensile strength (star and diamond) are derived from the values presented in Figure 4 of \citet{Skorov_2012}.}
    \label{fig:tensilestrength}
\end{figure}

The inner tensile strengths of four of the CoPhyLab mixtures plus the two pure materials were measured 10 times each. The average and standard deviations (small with regard to the values, nearly invisible in the Figure) for each sample are shown in Figure~\ref{fig:tensilestrength}. All the samples had varying porosities, in general between 65 and 75\,\%. As these values are close to those expected for the comet (70 to 80\,\%), we can use the values measured with only a small linear correction \citep[see][for details]{Bischoff2020}. Although not directly comparable, Figure~\ref{fig:tensilestrength} also displays the cometary values of the inter tensile strength in the greyed area (equivalent to 67P/C-G tensile strength between 0.02 and 15\,Pa,). The inner tensile strengths of the candidate analogues are higher by at least one order of magnitude compared to the inter tensile strengths derived for 67P/C-G which is to be expected as agglomerates are not as strongly bounded together when compared to grains. If we correct for temperature, we would expect a higher inner tensile strength for the candidates and if we correct for porosity, we should see a slight drop. It has been shown by experimental studies that comets could indeed be made of agglomerates with an inner agglomerate tensile strength on the order of 1\,kPa and an agglomerate tensile strength in the Pa range (\citealt{BLUM2014b,Gundlach2018}, Figure 7 in \citealt{Groussin2019}, and references therein).

Using the measured tensile strengths, we can try and limit the possible candidates. If we assume that we need an internal tensile strength of around 1\,kPa at approximately 70\,\% porosity, then all samples with a silica dust mass fraction higher than 0.7 would have the required tensile strength. However, as our measurements were done at room temperature -– and at cometary temperatures the tensile strength is expected to change (this will be investigated in the future) –- the candidates with a silica dust mass ratio below 70\,\% cannot be completely discarded.

\subsection{Thermal properties}

The evolution of the nucleus, from changes in morphology to the observed activity, is strongly linked to the thermal properties of cometary dust. Both modelling and instrument data analysis would greatly benefit from a better understanding of these quantities. The three main properties of interest are the thermal conductivity $k$, the heat capacity at constant pressure $C_p$, and the thermal inertia $I$, the latter defined through
\begin{equation}
I = \sqrt{k C_p \rho},
\end{equation}
with $\rho$ being the mass density of the sample.

These three parameters have a strong dependency on composition, porosity and temperature. Often their mixing laws are not linear. In-depth measurements of these properties at low temperatures, relevant to cometary environments, is out of the scope of this paper. Instead, an overview of these properties at room temperature is presented here, in order to attempt to constrain the best candidates.
 
\subsubsection{Cometary thermal properties}
\label{sec:knownthermalproperties}

Three instruments were able to assess the thermal properties of the cometary material. The MUPUS penetrometer thermal mapper was equipped to measure the thermal inertia of the upper layers. A local thermal inertia of 85\mmp 35\ti was required to reproduce the MUPUS-TM measurements \citet{Spohn2015}. 
Using data from VIRTIS, \citet{Leyrat2016} found that the best-fit thermal inertia for certain regions of 67P/CG is around 25-170\ti. Additionally, \cite{Marshall2018} fit a thermal model to the VIRTIS observations and found a thermal inertia of 80\ti. The Microwave Instrument for the Rosetta Orbiter (MIRO) has also helped to constrain the thermal properties of the upper subsurface. \citet{Gulkis2015} found a low thermal inertia of 10 to 50\ti by analysing the surface radiation emitted from the nucleus. Later, slightly different best-fit values to the MIRO measurements, equal to 10-30\ti and 10-60\ti, were calculated \citep{Choukroun2015, Schloerb2015}.

Another approach to constrain the thermal conductivity of cometary analogue dust candidates is to examine the thermal conductivity of carbonaceous chrondrites, which could be good analogues to the cometary refractory material. \citet{Henke2016} modelled the thermal conductivity of porous chondritic material based on the properties of the individual mixture components. They were able to derive a formula for the porosity dependence of the thermal conductivity
\begin{equation}
k(\phi) = ((k_b \mathrm{max}((1-2.216\phi),0))^4 + (k_b e^{1.2-\phi/0.167})^4)^{1/4},
\end{equation}
where $\phi$ is the porosity and $k_b$ the bulk thermal conductivity at 0\,\% porosity. Using this formula, we find a thermal conductivity between 0.05 and 0.2\tc for porosities between 70\,\% and 80\,\% and a bulk heat conductivity between 4.0 and 4.9\tc. 

\cite{Grott2019} used two formulas to derive the thermal conductivity of a boulder located on asteroid Ryugu (observed by the Hayabusa 2 spacecraft). These formulas are based on fits to experimental data of the thermal conductivity of carbonaceous chrondrites and read
\begin{equation}
k(\phi) = \frac{0.11(1-\phi)}{\phi}
\end{equation}
and 
\begin{equation}
k(\phi) = 4.3e^{-\phi/0.08}.
\end{equation}
With these formulas, we find a thermal conductivity between 6.8\,10$^{-4}$ and 3.6\,10$^{-2}$\tc for porosities between 70\,\% and 80\,\%. 

\subsubsection{Thermal properties measurement method}

The thermal properties of our dust candidates were measured using a KD2 Pro thermal Analyser. This device has the ability to measure both the thermal conductivity and the heat capacity of samples. We produced compressed cylinders with 2.5\,cm radius and varying heights (between 6 and 10\,cm) and inserted the device's 6\,cm needle inside the samples. For each mixture, we produced six samples with slightly varying porosities ranging from 60 to 75\,\%. 

\subsubsection{CoPhyLab dust thermal properties}

The measured thermal properties are for samples with a porosity in the range 60 to 75\,\%. In order to compare the measured values to the estimated cometary values described previously, we would need to know these properties in the porosity range 70 to 80\,\%. Making samples with porosities larger than 75\,\% is difficult and not possible with the current setup. Therefore we decided to fit the thermal properties with a linear fit and then extrapolate the measured values from the sample porosities (60 to to 75\,\%) to the cometary porosities (70 to 80\,\%). We make the assumption that this extrapolation is possible given that the porosities measured for the comet are very close to those of our samples. Additionally, though it would help to derive a more accurate mixing law, we lack measurements over a larger porosity range.

\begin{figure}
	\includegraphics[width=\columnwidth]{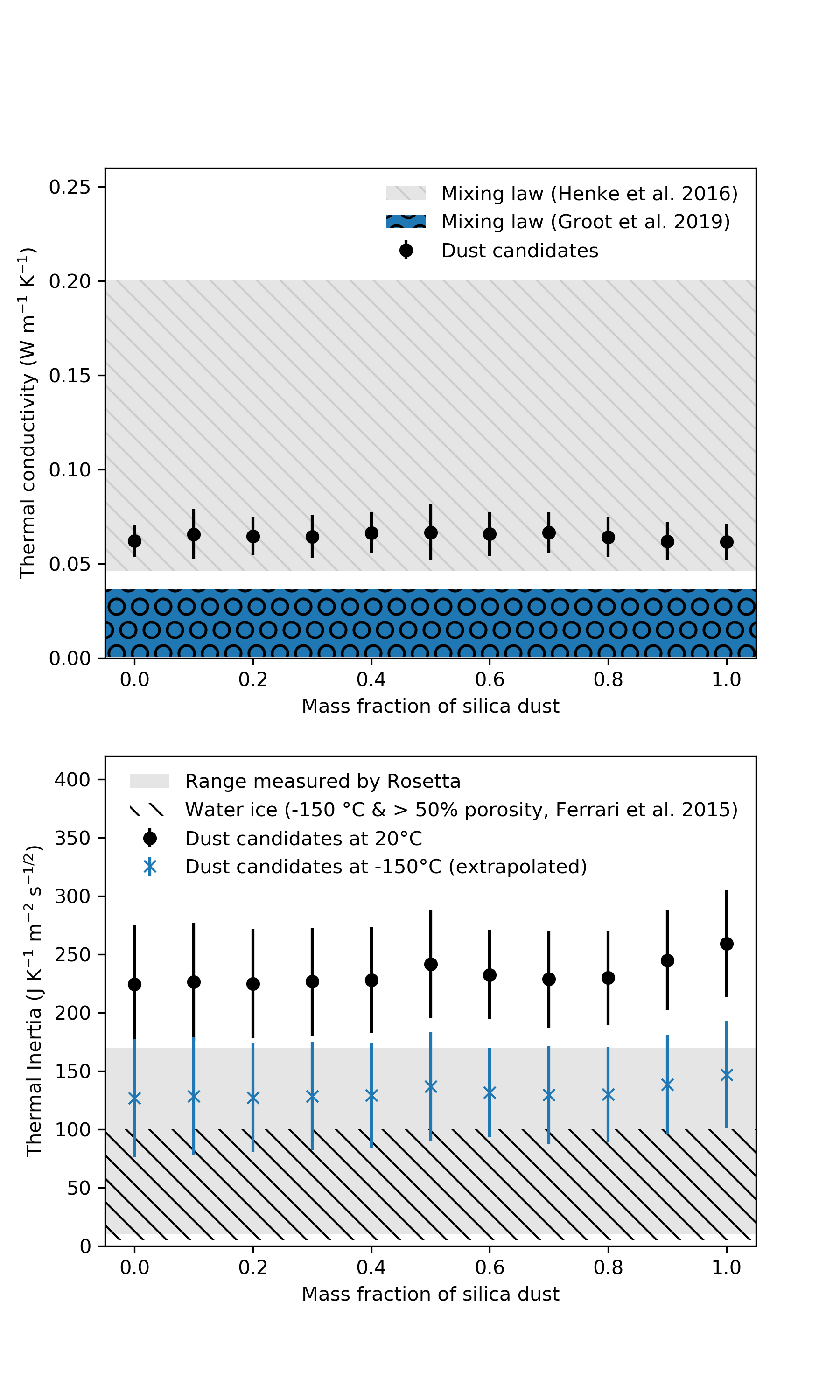}
    \caption{Top: Heat conductivity of the CoPhyLab cometary dust analogues as a function of the silica dust content. The two grey-shaded areas represent the thermal conductivity of carbonaceous chrondrites with a porosity between 70-80\,\% based on mixing models. Bottom: Thermal inertia of the CoPhyLab cometary dust analogues as a function of the silica dust content. The grey-shaded area represents the thermal inertia values constrained by Rosetta described in the text.}
    \label{fig:ThermalFig}
\end{figure}

The measured properties extrapolated to a porosity of 70-80\,\% are shown in Figure~\ref{fig:ThermalFig}. The top figure shows the thermal conductivity as a function of the silica dust mass fraction in the mixture. We also display the ranges of possible porous chondritic mixtures determined by the mixing laws presented in Section~\ref{sec:knownthermalproperties}, for a porosity of 70-80\,\%.

We observe that the thermal conductivity at such high porosities has no visible dependencies on the silica dust ratio. This is most likely due to the fact that the porosity of the sample is the main factor determining the thermal properties for two main reasons. (i) The measurements were done at atmospheric pressure. The high porosity of our samples means that air contributes strongly to the global thermal conductivity of our sample (see below). (ii) Radiation between particles in the sample also contributes to the global thermal conductivity, particularly for high porosities. All the candidates have a thermal conductivity in the ranges determined by \citet{Henke2016} and slightly higher than \citet{Grott2019}. 

The bottom plot of Figure~\ref{fig:ThermalFig} shows the thermal inertia of the samples as a function of the silica dust content. The grey-shaded area represents the range of thermal inertia measured on comet 67P/C-G by the Rosetta instruments. The measured thermal inertia for our samples is higher by a factor of 1.5 - 2 and does not overlap with the cometary range. However, our analogue-material values are generally overestimated for three main reasons. (i) The measurements were not performed in vacuum due to experimental constraints (the measurement device is not adapted to work in a vacuum environment) and therefore the voids in our samples were filled with air. In a vacuum environment, this would tend to reduce the thermal conductivity of our candidates, as air has a thermal conductivity of k = 0.026\tc~at 298\,K. (ii) The thermal properties were measured at room temperature and, in general, geological materials will see their thermal inertia reduced as their temperature drops. This can be seen for example in Figure 10 of \citet{Opeil2020} for carbonaceous chondrites, in which the thermal inertia drops by 15-20\,\% when the temperature drops from 300\,K to 150\,K. By applying the same thermal inertia change to our samples, we obtain the blue crosses in Figure~\ref{fig:ThermalFig}. This puts our measured values in the range of the values measured by Rosetta. (iii) One additional change that impacts the thermal properties is the presence of water ice. Rosetta determined that the presence of water ice is likely in the areas studied and granular water ice has a thermal inertia between 10 and 100\ti~at low temperatures and high porosities \citep{Ferrari2016}. The hatched area in Figure~\ref{fig:ThermalFig} denotes this range. Therefore, we interpret that any mixture of our candidates and water ice would probably have a thermal inertia well in the range measured by the Rosetta instruments.

The thermal properties are thus constrained by porosity, temperature, and the presence of volatiles. An in-depth study of these parameters (in vacuum and mixed with water ice at low temperature) for the chosen CoPhyLab dust is planned, but is out of the scope of this paper, as it most likely will have little to no impact on the selection of an appropriate candidate for a cometary physical analogue, as the porosity and temperature are well controlled properties for our samples.

\subsection{Electrical properties}

The electrical properties of a material are described by its relative electric permittivity. This number is a complex, dimensionless factor 
\begin{equation}
\epsilon_r = \epsilon_\mathrm{r}' -i\epsilon_\mathrm{r}'' 
\end{equation}
that measures how an electric field decreases relative to vacuum when it passes through a material.

The real part $\epsilon_\mathrm{r}'$ (also known as the dielectric constant) represents the capacity of the material to change polarisation as an imposed electric field changes (i.e. it relates to energy storage), while the imaginary part $\epsilon_\mathrm{r}''$ represents the ability of the material to transmit the electrical charge through charged particles (i.e. it relates to energy attenuation). In practice, an electromagnetic wave changes its velocity (depending on $\epsilon_\mathrm{r}'$) and magnitude (depending on the loss tangent $\tan(\delta) = \epsilon_\mathrm{r}''/\epsilon_\mathrm{r}'$) as it travels through a material. Thus, instruments that can measure these changes in the wave can provide direct information about the permittivity of a material. 

Both the real and imaginary parts of the complex permittivity depend strongly on the composition, porosity, and temperature of the material. This makes permittivity an extremely useful physical property in the characterisation of planetary surfaces. There are many remote and in-situ instruments that can measure the permittivity of a planetary surface material with varying degrees of resolutions in space and frequency. Most of these instruments operate in the microwave frequency range, where the dependence of the permittivity on frequency is weaker than at other spectral ranges. For example, by increasing order of frequency ranges, permittivity probes, echo-sounding radars, surface imaging radars, and microwave radiometers are all able to constrain the complex permittivity of a planetary surface.

\subsubsection{Electrical properties of comets}

The electrical properties of cometary matter have been studied by earth-based telescopes, orbiting spacecrafts, and in-situ landed instruments. \citet{Kamoun2014} used the Arecibo radio-telescope to constrain the real part of the permittivity of the top 2.5 meters of comet 67P/C-G to be between 1.9 - 2.1 at a frequency of 2.38\,GHz. Initial measurements by the CONSERT bistatic radar on Rosetta estimated $\epsilon_\mathrm{r}'=1.27$ at 90\,MHz, valid for depths up to 100\,m \citep{Kofman2015}. However, with the subsequent determination of the exact location of the Philae lander (that carried one of the CONSERT antennas, cf. \citealt{Kofman2007}), $\epsilon_\mathrm{r}'$ was found to range between 1.7 and 1.95 in the shallow subsurface of 67P/C-G (< 25\,m) and between 1.2 and 1.32 in the deeper interior (25 – 150\,m  \citealt{Kofman2020}), in better agreement with \citet{Kamoun2014}. Finally, data collected at 409\,Hz by the SESAME-PP probe on board Philae led to values of $\epsilon_\mathrm{r}'$ of no less than 2.45 for the top few meters of the surface beneath the landing site \citep{Lethuillier2016}. The $\epsilon_\mathrm{r}'$ of most geologic materials lose their frequency dependency at the low temperatures relevant to cometary environments \citep{Brouet2016}. Thus, the difference observed in the values of $\epsilon_\mathrm{r}'$ at high (Arecibo and CONSERT) and low (SESAME-PP) frequencies is most likely due to the presence of a porosity gradient inside the nucleus of the comet \citep{Lethuillier2016}.

The imaginary part of the permittivity is difficult to estimate from these measurements, due not only to technical constraints, but also to the fact that it has a much stronger temperature and frequency dependency than the real part, complicating its prediction outside of the measured range. Nevertheless, based on the materials that we expect to be present (e.g. ices) and its probable high porosity, attenuation should be low within the nucleus and thus $\epsilon_\mathrm{r}''$ is also likely to be low \citep{Kofman2015}.


\subsubsection{Electrical properties measurement method}

The laboratory setup (Epsimu (R) from Multiwave) we used for the electrical characterisation of our CoPhyLab dust mixtures consists of a conical coaxial cell connected to a 2-port Vector Network Analyzer (VNA; Anritsu Master MS2038C) equipped with a central cylindrical sample holder, which we filled with the various dust simulant samples. The samples were contained by two dielectric walls of polytetrafluoroethylene (PTFE). We used this system \citep{Sabouroux2011} to measure the permittivity of the samples between 80 MHz and 3\,GHz. This choice of frequency range does not include the lower frequencies at which SESAME-PP operated, as the measurement system would fail to operate appropriately at such long wavelengths. However, the higher frequency values, rather than the lower ones, were those used to probe deep into the nucleus of 67P/C-G and are thus more relevant to this study, since they are characteristic of the entire comet. They also have the advantage of having a higher accuracy (the SESAME-PP values are only an upper limit to the real part of the electrical properties). All measurements were made at ambient temperature (23$^{\circ}$\,C) and under ambient air pressure. A full description of our experimental setup can be found in \citet{Brouet2015}, and further examples of its use for planetary regolith studies can be found in \citet{Brouet2016,Brouet2019}. 

The sample preparation consists of completely filling the sample container with the sample. In order to achieve different porosities within the sample holder, the sample was compressed throughout the filling process with varying degrees of pressure, using a cylinder cut precisely to fit the dimensions of the sample holder. Once filled, the sample holder was weighed to determine the mass and bulk density of the sample, using the known volume of the sample containment volume (1.291 ± 0.288\,cm$^3$). Using the grain densities determined for the components and their mixtures (see Section \ref{sec:materialproperties}), we calculated the porosities of each sample measured with this setup. We limited our measurements to two porosities for each dust mixture, one high (70\mmp7\,\%) and one low (59\mmp3\,\%). Lower porosities are difficult to achieve for dry granular samples, but these relatively high porosities are more relevant to the observed porous nature of comets. 

The permittivity is obtained by first measuring the scattering parameters (the reflection and transmission coefficients) of the cell loaded with the relevant sample over 1000 frequencies between 3\,MHz and 3\,GHz. The permittivity of the sample at each frequency is then determined with the Nicolson-Ross-Weir procedure \citep{Nicolson1970} applied to the sample scattering parameters. As there is little variability in permittivity in the frequency range between 80\,MHz and 3\,GHz, we calculated the average permittivity in this frequency range for each sample, and took this to be the representative value of the sample. Further details on the calculation procedure and its use with this setup can be found in \citet{Georget2014}.

\subsubsection{Electrical properties of CoPhyLab dust}

\begin{figure}
	\includegraphics[width=\columnwidth]{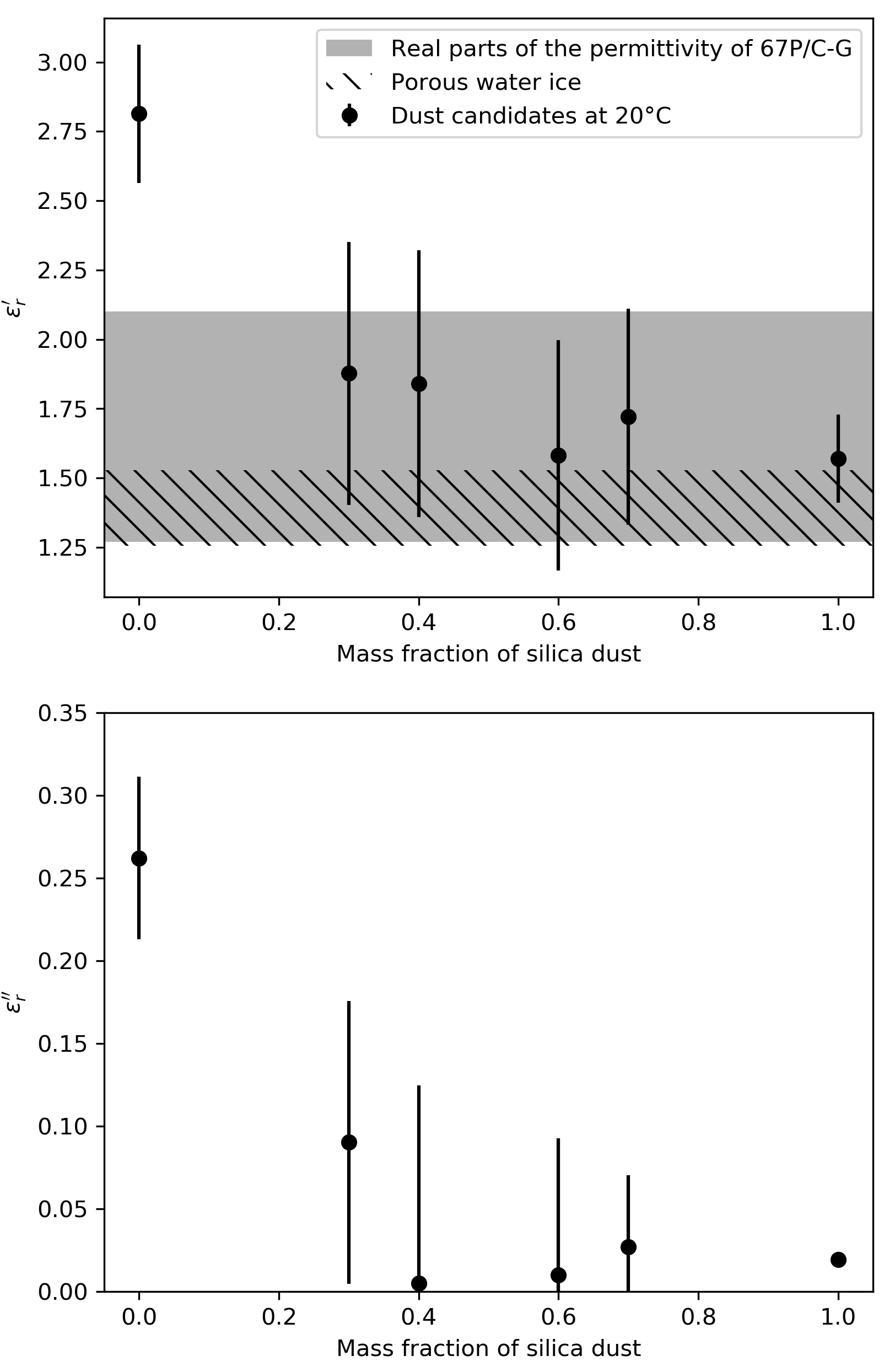}
    \caption{Top: Real part of the permittivity of porous (70 – 85\,\%) CoPhyLab dust candidates as a function of the silica dust content. The grey-shaded area represents the range of values observed for the nucleus of 67P/C-G (\citealt{Kamoun2014,Kofman2015}) and the hatched area represents the range of $\epsilon_\mathrm{r'}$ for cold, porous (70-85\,\%) water ice. The error bars take into account both the uncertainty of the measurements combined with the range of porosities that we extrapolated to (70 to 80\,\%).
    Bottom: Imaginary part of the permittivity of porous (70 – 85\,\%) CoPhyLab dust candidates as a function of the silica dust content.}
    \label{fig:PermittivityFig}
\end{figure}

The measured real and imaginary parts of the permittivity of the end member components and selected mixtures are shown in Figure~\ref{fig:PermittivityFig}. In order to obtain these values, we took the results found for both the high  (70\mmp 7\,\%) and low (59\mmp 3\,\%) porosities, then performed a linear fit to obtain the porosity dependence of the electrical properties. Using this linear relation, we extrapolated the electrical properties for the porosity of the nucleus of 67P/C-G (70 to 80\,\%). The error bars represent both the uncertainty of the measurements and the range of porosities to which we extrapolated to (70 to 80\,\%). Many mixing laws exist that are used to extrapolate the electrical permittivity of granular samples \citep[e.g.][]{Sihvola2000, Brouet2016}. However, given the low measured permittivities and, importantly, given that the cometary porosities are close to those measured for the lab samples, we decided to extrapolate linearly. A future in-depth study of the electrical properties of the CoPhyLab dust will investigate the possible mixing laws to be used and will compare these with an empirical law devised specifically for the CoPhyLab dust components.

Figure~\ref{fig:PermittivityFig} shows that all samples with a silica dust content higher than 30\,\% by mass have values of $\epsilon_\mathrm{r'}$ compatible with those measured for 67P/C-G (grey-shaded area). As cometary matter is assumed to be a mixture of water ice and dust, the permittivity measured at 67P/C-G must include the contribution of water ice. Thus, we also indicated in Figure~\ref{fig:PermittivityFig} the theoretical $\epsilon_\mathrm{r'}$ of cold (150\,K) and porous (between 70 and 80\,\%) water ice, which ranges between 1.25 and 1.50 (based on the Debye model, \citealt{Debye1929}). Any mixture of this water ice with CoPhyLab dust of more than 30\,\% silica dust by mass would still be compatible with the measured cometary values. We therefore consider all dust analogues with more than 30\,\% silica by mass to be valid from the viewpoint of their electrical properties.

\section{Cophylab dust}
\label{sec:cophylabdust}

Using the constraints on all the physical properties described throughout this paper, it is possible to build a physical analogue for cometary dust. In this Section, we describe the method we used to determine this analogue and summarise the physical properties of the chosen mixture.

\subsection{Choosing the best ratio}

Table~\ref{table:prosConsOptionalApproaches} shows all of the physical properties described above for each mass ratio between silica and charcoal dust. The values are indicated with the target necessary to build a plausible cometary non-volatile physical analogue. Values indicated in green are optimal, values in orange are less desirable, and values in pink are to be avoided. For each mixture, we associated a numerical value to the quality of the fitting (1 for green, 0.5 for orange, and 0 for pink) and when adding up these values, we obtained a total compatibility for this mixture. This implies that the optimal mixing ratio for a dust analogue to plausibly simulate cometary regolith in physical experiments is 60/40\,\% or 70/30\,\% of silica dust/charcoal by mass.

\begin{table*}
\caption{Physical properties of all investigated dust analogue candidates with target ranges indicated in the second column. The cells in green represent values that fall within the constraints; the orange cells have values that fall outside the desired ranges, but could still be acceptable, and the pink cells denote unacceptable values. For each mixture, we associated a numerical value to the quality of the fitting (1 for green, 0.5 for orange, and 0 for pink) and when adding up these values, we obtained a total compatibility for this mixture (bottom row). The mixtures with a 60/40\,\% and 70/30\,\% SiO$_2$/charcoal both have the highest total compatibility and therefore represent the best candidate dust analogue.}
\label{table:prosConsOptionalApproaches}
\begin{tabular}{llccccccccccc}
\hline
Physical property & Target values & 0/100 & 10/90 & 20/80& 30/70& 40/60& 50/50& 60/40& 70/30& 80/20& 90/10 & 100/0 \\

\hline
Grain size & $<$ 1\,\micron & \multicolumn{11}{c}{\cellcolor{blu}$<$ 1\,\micron}\\ \arrayrulecolor{gray}\cdashline{1-2}
Agglomerates & \makecell{Power law index:\\-1.8 to -2.3} & NM &\cellcolor{rei} -1.05 & \cellcolor{blu} -1.81 & \cellcolor{blu} -1.86 & \cellcolor{blu} -1.97 & \cellcolor{blu} -2.22 & \cellcolor{blu} -2.26 & \cellcolor{blu} -2.23 & \cellcolor{blu} -1.83 & \cellcolor{blu} -1.79 & NM\\\arrayrulecolor{gray}\cdashline{1-2}
Density & 2-3\,g\,cm$^{-3}$ &\cellcolor{rei} 1.55 &\cellcolor{rei} 1.65 &\cellcolor{rei} 1.75 &\cellcolor{rei} 1.86 &\cellcolor{ver} 1.96 &\cellcolor{blu} 2.06 &\cellcolor{blu} 2.17
&\cellcolor{blu} 2.27 &\cellcolor{blu} 2.37 &\cellcolor{blu} 2.47 &\cellcolor{blu} 2.58\\\arrayrulecolor{gray}\cdashline{1-2}
\makecell[l]{Albedo\\(550 nm)} & 1.47 & \cellcolor{ver} 3.00  & \cellcolor{ver} 3.08  & \cellcolor{ver} 3.38  & \cellcolor{ver} 3.86  & \cellcolor{ver} 3.79  & \cellcolor{ver} 4.43  & \cellcolor{ver} 3.47  & \cellcolor{rei} 6.48 & \cellcolor{rei} 7.52 & \cellcolor{rei} 25.26 & \cellcolor{rei} 85.12\\\arrayrulecolor{gray}\cdashline{1-2}
Thermal inertia & \makecell{10 - 170\\J\,m$^{-2}$\,K$^{-1}$\,s$^{-1/2}$} & \cellcolor{blu} 127.0 & \cellcolor{blu} 128.2  & \cellcolor{blu} 127.2  & \cellcolor{blu} 128.3  & \cellcolor{blu} 129.1  & \cellcolor{blu} 136.8  & \cellcolor{blu} 131.6  & \cellcolor{blu} 129.4  & \cellcolor{blu} 130.0  & \cellcolor{blu} 138.5  & \cellcolor{blu} 146.8 \\\arrayrulecolor{gray}\cdashline{1-2}

\makecell[l]{Inter Tensile \\strength} & $>$1\,kPa  & \cellcolor{ver} 109 &NM &NM &NM &NM & \cellcolor{ver} 309 & \cellcolor{ver} 551 & \cellcolor{blu} 932 & \cellcolor{blu} 1427 &NM&  \cellcolor{blu} 8667 \\\arrayrulecolor{gray}\cdashline{1-2}
\makecell[l]{Electric permittivity\\(real part)} & 1.27-2.1 & \cellcolor{ver} 2.81 & NM & NM &  \cellcolor{blu} 1.88&  \cellcolor{blu} 1.84 & NM &  \cellcolor{blu} 1.58 &  \cellcolor{blu} 1.72 &NM & NM &  \cellcolor{blu} 1.57 \\

\hline

Total compatibility &&3.5&2.5&3.5&4.5&5&5&6&6&5&4&5\\

\hline
\end{tabular}
\end{table*}

\subsection{Dust recipe}
\label{mixingrecipie}

In order to create a visibly homogeneous dust in which both components are intimately mixed, we used the following recipe:

\begin{enumerate}
\item Weigh 60\,\% (or 70\,\%) of the total desired mass in silica dust and 40\,\% (or 30\,\%) of it in charcoal.
\item Roughly mix both components in a container.
\item Pour a chosen liquid in another container (e.g., a nitrogen dewar if using liquid nitrogen as the mixing liquid). This step in necessary to dissolve previously formed agglomerates.
\item Spoon the mixed dust into the liquid while stirring.
\item Once all the dust is in the liquid, continue stirring until all agglomerates are broken down and both components are well mixed.
\item Let the mixture air dry. Some frost may form but can removed by putting the dust in a desiccator.
\item Shake the mixture with an automatic shaker or by hand for 10 minutes to form new agglomerates formed of grains from the two components.
\end{enumerate}

Distilled water, ethanol or liquid nitrogen can all be used as the mixing liquid. However, this will affect the drying time and handling of the dust. In all cases, a dust-protection mask is required to avoid breathing in the smallest dust particles and gloves are suggested as the charcoal can be messy when handled. Additional caution must be taken when liquid nitrogen is used as it will disperse the dust in the air more easily than water or ethanol, especially during step (iv). Finally, the handling should be done in an air-suction hood (first because of the powder and second because of the evaporating nitrogen that displaces the air)

\begin{figure}
	\includegraphics[width=\columnwidth]{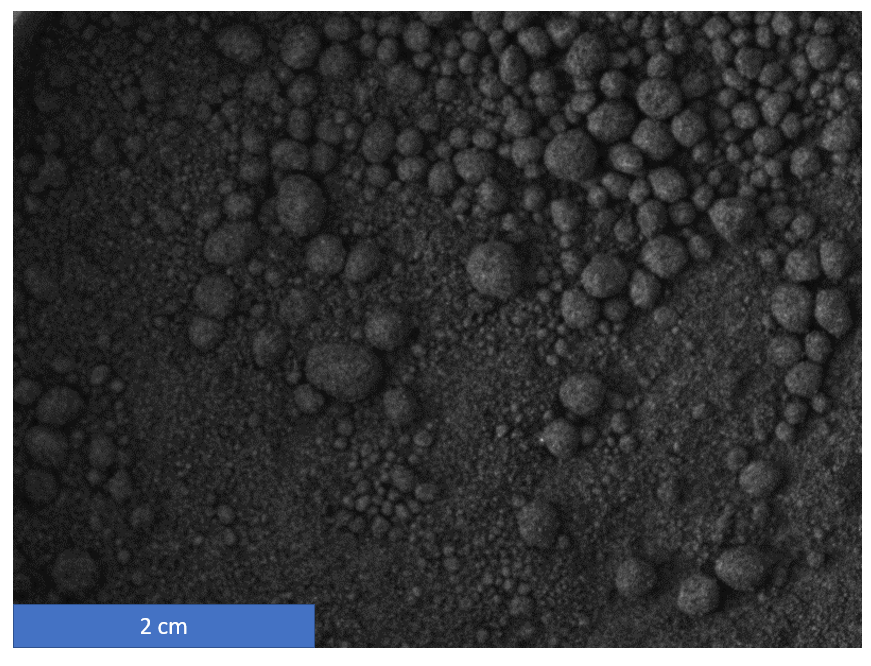}
    \caption{Image of the CoPhyLab cometary dust analogue.}
    \label{fig:dustimage}
\end{figure}

An image of the resulting dust is shown in Figure~\ref{fig:dustimage}. Its low reflectance and the presence of large agglomerates surrounded by smaller aggregates and grains are visible. Samples only composed of agglomerates can be obtained by sieving gently the CoPhyLab dust.

\subsection{Physical properties}

Table~\ref{table:physicalpropertiesfinal} summarises the measured physical properties of our chosen dust analogues. They mostly match the constraints presented in Section~\ref{sec:materialproperties}. Both candidates have an albedo higher than that of comets, but the 60/40\,\% has the lowest value out of both (3.47 compared to 1.47 for 67P/C-G). We therefore decided to use slightly higher albedo values, as this should not impact the results in a significant way given that the value is well characterised and can be modelled in numerical simulations.

The inner tensile strength of the 60/40\,\% mixture is lower than the value aimed for  (551\,Pa compared to around 1\,kPa (see Section~\ref{fig:tensilestrength}), whereas the 70/30\,\% has a good match (932\,Pa). The inner tensile strength is not a strong constraint, because the inter tensile strength is hard to measure for this dust analogue \citep[see][for details]{BLUM2014b} and the inner tensile strength of the comet is not known.

It is important to note here that one of the biggest facing in cometary experiments trying to reproduce the activity is the presence of earth gravity.  One possibility to overcome this limitation is to use a micro-g environment as given in a drop tower or during a parabolic flight. But both is beyond the scope of the CoPhyLab arrangements. Another option to adapt the experiments to gravity on Earth is the scaling of the guiding force ratios - gravity/cohesion and gravity/gas drag - by reducing the grain size distribution compared to a real comet and by increasing the strength of sample illumination. However, we are limited in these  aspects as reducing the grain sizes is quite difficult and a strong radiation source, as would be needed, is currently not available . Upcoming papers using the CoPhyLab dust will discuss in much more depth the gravity issue.

For most of the other physical properties, both dust candidates are very similar. The choice of which ratio would be better for a cometary physics dust analogue depends on which experiments are planned. In general, we favour the 60/40\,\% for all experiments with the exception of gas flow measurements (the 70/30\,\% was easier to handle for these experiments) and for tensile strength measurements (the 70/30\,\% is a better match).

\begin{table}
\caption{The physical properties of the best fitting dust analogues.}
\label{table:physicalpropertiesfinal}
\centering
\begin{tabular}{l|c|c}
\hline
Physical property & 60/40\,\% & 70/30\,\%\\
\hline
Average grain size & 392\,nm  & 364\,nm\\
\makecell[l]{Agglomerate size\\power law index} & -2.12 & -2.23\\
Density & 2.17\,g/cm$^3$& 2.27\,g/cm$^3$ \\
Albedo (550\,nm) & 3.47 & 6.48\\
535-880\,nm slope & 4.89& 4.89\\
Gas permeability & N/A & 2.0415e-11\,m\\
\makecell[l]{Knudsen diffusion\\coefficient} &  N/A & 1.7366e-03\,m$^2$/s\\
\makecell[l]{Thermal inertia\\(J/m$^2$s$^{1/2}$K)} & 129.4 & 131.6\\
Tensile strength & 551\,Pa & 923\,Pa\\
\makecell[l]{Electric Permittivity\\(real part)} &  1.58\mmp 0.4 & 1.75\mmp 0.4\\
\makecell[l]{Electric Permittivity\\(imaginary part)} & 0.011\mmp 0.08 & 0.023\mmp 0.08 \\
\hline
\end{tabular}
\end{table}

\section{Conclusions}
In this paper, we described in detail the measured physical parameters of the non-volatile part of cometary dust. We then compared these values to the physical properties of cometary dust analogues with the objective of finding the analogue that would allow us to study the physics of comets in the most accurate way possible. For this, we made a dust analogue composed of silica dust to represent the siliceous material present in cometary dust and ground charcoal to represent the organic and dark material observed on the nucleus of comets. For most properties, we produced samples ranging from 10/90\,\% to 90/10\,\% in mass ratio of silica dust/charcoal and compared their properties to the measured cometary values. Once this was done, we attributed a match descriptor (1 for a good match, 0.5 for an average match and 0 for a poor match) and were able to concluded that a ratio of either 60/40\,\% or 70/30\,\% of silica dust/charcoal represents good analogues for the physical properties of the non-volatile part of cometary material. We generally favour the 60/40\,\% ratio, because it has a better matching albedo, which is an important parameter that controls how much of the insolation power is absorbed.
 
We also presented the laboratory protocol we used in order to produce the CoPhyLab dust as this should allow the scientific community to make use of it as well.
 
In conclusion, we find that the CoPhyLab dust is a good analogue that will be used in many future experiments planned for the study of the Physics of comets.

\section*{Acknowledgements}
This work was carried out in the framework of the CoPhyLab project funded by the D-A-CH programme (DFG: GU 1620/3-1 and BL 298/26-1 / SNF: 200021E 177964 / FWF: I 3730-N36). The team from the University of Bern is supported by the Swiss National National Science Foundation, in part through the NCCR PlanetS

\section*{Data Availability}

The data for Sections 3.2, 3.4, 3.5, 3.6, 3.7 and 3.8 are available in the article. The data for Sections 3.1 and 3.3 will be shared on request to the corresponding author.



\bibliographystyle{mnras}
\bibliography{dustPaper} 




\appendix


\bsp	
\label{lastpage}
\end{document}